\documentclass [twocolumn]{aastex631}
\usepackage{hyperref} 
\usepackage{amssymb, amsmath, mathtools}
\usepackage{natbib}
\usepackage{enumerate}
\usepackage{graphicx}

\usepackage{latexsym}
\usepackage{color}
\usepackage{graphicx}
\usepackage{epstopdf}
\usepackage{threeparttable}
\usepackage[T1]{fontenc}
\usepackage{tabularx}
\usepackage{multirow}
\usepackage{colortbl}
\usepackage[version=4]{mhchem} 
\usepackage{xspace}

\DeclareSymbolFont{UPM}{U}{eur}{m}{n}
\DeclareMathSymbol{\umu}{0}{UPM}{"16}

\definecolor{xkcdbrightblue}{RGB}{43, 103, 252}
\newcommand{\highlight}[1]{{\color{magenta} \bf #1}} 

\newcommand\degree{\degr}
\newcommand\degrees\degree

\DeclareSymbolFont{UPM}{U}{eur}{m}{n}
\DeclareMathSymbol{\umu}{0}{UPM}{"16}
\let\oldumu=\umu
\renewcommand\umu{\ifmmode\oldumu\else\math{\oldumu}\fi}

\newcommand\microns \micron

\let\oldsim=\sim
\renewcommand\sim{\ifmmode\oldsim\else\math{\oldsim}\fi}
\let\oldpm=\pm
\renewcommand\pm{\ifmmode\oldpm\else\math{\oldpm}\fi}
\newcommand\by{\ifmmode\times\else\math{\times}\fi}

\newbox{\wdbox}
\renewcommand\c{\setbox\wdbox=\hbox{,}\hspace{\wd\wdbox}}
\renewcommand\i{\setbox\wdbox=\hbox{i}\hspace{\wd\wdbox}}
\newcommand\n{\hspace{0.5em}}

\newcount\timect
\newcount\hourct
\newcount\minct
\newcommand\now{\timect=\time \divide\timect by 60
         \hourct=\timect \multiply\hourct by 60
         \minct=\time \advance\minct by -\hourct
         \number\timect:\ifnum \minct < 10 0\fi\number\minct}

\catcode`@=11

\newcommand\comment[1]{}
\newcommand\commenton{\catcode`\%=14}
\newcommand\commentoff{\catcode`\%=12}

\renewcommand\math[1]{$#1$}
\newcommand\mathshifton{\catcode`\$=3}
\newcommand\mathshiftoff{\catcode`\$=12}

\comment{the backslash is necessary}

\comment{alignment tab}

\let\atab=&
\newcommand\atabon{\catcode`\&=4}
\newcommand\ataboff{\catcode`\&=12}

\let\oldmsp=\sp
\let\oldmsb=\sb
\def\sp#1{\ifmmode
           \oldmsp{#1}%
         \else\strut\raise.85ex\hbox{\scriptsize #1}\fi}
\def\sb#1{\ifmmode
           \oldmsb{#1}%
         \else\strut\raise-.54ex\hbox{\scriptsize #1}\fi}
\newbox\@sp
\newbox\@sb
\def\sbp#1#2{\ifmmode%
           \oldmsb{#1}\oldmsp{#2}%
         \else
           \setbox\@sb=\hbox{\sb{#1}}%
           \setbox\@sp=\hbox{\sp{#2}}%
           \rlap{\copy\@sb}\copy\@sp
           \ifdim \wd\@sb >\wd\@sp
             \hskip -\wd\@sp \hskip \wd\@sb
           \fi
        \fi}
\def\msp#1{\ifmmode
           \oldmsp{#1}
         \else \math{\oldmsp{#1}}\fi}
\def\msb#1{\ifmmode
           \oldmsb{#1}
         \else \math{\oldmsb{#1}}\fi}
\def\supon{\catcode`\^=7}
\def\supoff{\catcode`\^=12}
\def\subon{\catcode`\_=8}
\def\suboff{\catcode`\_=12}
\def\supsubon{\supon \subon}
\def\supsuboff{\supoff \suboff}

\newcommand\actcharon{\catcode`\~=13}
\newcommand\actcharoff{\catcode`\~=12}

\newcommand\paramon{\catcode`\#=6}
\newcommand\paramoff{\catcode`\#=12}

\comment{And now to turn us totally on and off...}

\newcommand\reservedcharson{\commenton \mathshifton \atabon \supsubon \actcharon
	\paramon}

\newcommand\reservedcharsoff{\commentoff \mathshiftoff \ataboff
	\supsuboff \actcharoff \paramoff}

\catcode`@=12
\reservedcharsoff

\reservedcharson

\comment{ Must have ONLY ONE of these... trust these macros, they work

}

\newcommand{\squishlist}{
 \begin{list}{$\bullet$}
  { \setlength{\itemsep}{0pt}
     \setlength{\parsep}{0pt}
     \setlength{\topsep}{0pt}
     \setlength{\partopsep}{0pt}
     \setlength{\leftmargin}{2.0em}
     \setlength{\labelwidth}{1.5em}
     \setlength{\labelsep}{0.5em} } }

\newcommand{\squishlisttwo}{
 \begin{list}{$\bullet$}
  { \setlength{\itemsep}{1pt}
     \setlength{\parsep}{3pt}
     \setlength{\topsep}{3pt}
     \setlength{\partopsep}{0pt}
     \setlength{\leftmargin}{2.0em}
     \setlength{\labelwidth}{1.5em}
     \setlength{\labelsep}{0.5em} } }

\newcommand{\squishend}{
  \end{list}  }

\reservedcharson
\actcharon
\newcommand{\poseidon}{\texttt{POSEIDON}\xspace}
\newcommand{\eureka}{\texttt{Eureka!}\xspace}

\defcitealias{Lustig-YaegerFu2023}{Lustig-Yaeger \& Fu et al.}
\defcitealias{moranstevenson2023}{Moran \& Stevenson et al.}
\defcitealias{MayMacDonald2023}{May \& MacDonald et al.}
\defcitealias{CarterMay2024}{Carter \& May et al.}

\graphicspath{{./}{Figures/}}

\shorttitle{WASP-17b DREAMS of PIE}
\shortauthors{Lustig-Yaeger \& Sotzen et al.}

\begin{document}

\title{JWST-TST DREAMS: The Nightside Emission and Chemistry of WASP-17b}

\correspondingauthor{Jacob Lustig-Yaeger}
\email{Jacob.Lustig-Yaeger@jhuapl.edu}
\correspondingauthor{Kristin Showalter Sotzen}
\email{kristin.sotzen@jhuapl.edu, kshowal3@jhu.edu}


\author[0000-0002-0746-1980]{Jacob Lustig-Yaeger}
\affiliation{JHU Applied Physics Laboratory, 11100 Johns Hopkins Rd, Laurel, MD 20723, USA}

\author[0000-0001-7393-2368]{Kristin S. Sotzen}
\affiliation{JHU Applied Physics Laboratory, 11100 Johns Hopkins Rd, Laurel, MD 20723, USA}

\author[0000-0002-7352-7941]{Kevin B. Stevenson}
\affiliation{JHU Applied Physics Laboratory, 11100 Johns Hopkins Rd, Laurel, MD 20723, USA}

\author[0000-0002-8163-4608]{Shang-Min Tsai}
\affiliation{Department of Earth Sciences, University of California, Riverside, Riverside, CA, USA}
\affiliation{Institute of Astronomy \& Astrophysics, Academia Sinica, Taipei 10617, Taiwan} 


\author[0000-0002-8211-6538]{Ryan C. Challener}
\affiliation{Department of Astronomy and Carl Sagan Institute, Cornell University, 122 Sciences Drive, Ithaca, NY 14853, USA} 

\author[0000-0002-8515-7204]{Jayesh Goyal}
\affiliation{School of Earth and Planetary Sciences (SEPS), National Institute of Science Education and Research (NISER), HBNI, Odisha, India}

\author[0000-0002-8507-1304]{Nikole K. Lewis}
\affiliation{Department of Astronomy and Carl Sagan Institute, Cornell University, 122 Sciences Drive, Ithaca, NY 14853, USA}

\author[0000-0002-2457-272X]{Dana R. Louie}
\affiliation{Catholic University of America, Department of Physics, Washington, DC, 20064, USA}
\affiliation{Exoplanets and Stellar Astrophysics Laboratory (Code 667), NASA Goddard Space Flight Center, Greenbelt, MD 20771, USA}
\affiliation{Center for Research and Exploration in Space Science and Technology II, NASA/GSFC, Greenbelt, MD 20771, USA}

\author[0000-0002-4321-4581]{L. C. Mayorga} 
\affiliation{JHU Applied Physics Laboratory, 11100 Johns Hopkins Rd, Laurel, MD 20723, USA}

\author[0000-0002-2643-6836]{Daniel Valentine}
\affiliation{University of Bristol, HH Wills Physics Laboratory, Tyndall Avenue, Bristol, UK}

\author[0000-0003-4328-3867]{Hannah R. Wakeford}
\affiliation{University of Bristol, HH Wills Physics Laboratory, Tyndall Avenue, Bristol, UK} 


\author[0000-0001-8703-7751]{Lili Alderson} 
\affiliation{Department of Astronomy and Carl Sagan Institute, Cornell University, 122 Sciences Drive, Ithaca, NY 14853, USA} 

\author[0000-0002-0832-710X]{Natalie H. Allen}
\affiliation{William H. Miller III Department of Physics and Astronomy, Johns Hopkins University, Baltimore, MD 21218, USA}

\author[0000-0002-5967-9631]{Thomas J. Fauchez}
\affiliation{NASA Goddard Space Flight Center, 8800 Greenbelt Road, Greenbelt, MD 20771, USA}
\affiliation{Integrated Space Science and Technology Institute, Department of Physics, American University, Washington DC, USA}
\affiliation{NASA GSFC Sellers Exoplanet Environments Collaboration, USA}

\author[0000-0002-5322-2315]{Ana Glidden}
\affiliation{Department of Earth, Atmospheric and Planetary Sciences, Massachusetts Institute of Technology, Cambridge, MA 02139, USA}
\affiliation{Kavli Institute for Astrophysics and Space Research, Massachusetts Institute of Technology, Cambridge, MA 02139, USA}

\author[0000-0003-0854-3002]{Am\'{e}lie Gressier}
\affiliation{Space Telescope Science Institute, 3700 San Martin Drive, Baltimore, MD 21218, USA}

\author[0000-0003-4596-0702]{Sarah M. Hörst}
\affiliation{Johns Hopkins University Department of Earth and Planetary Sciences Baltimore MD 21218, USA}

\author[0000-0001-5732-8531]{Jingcheng Huang}
\affiliation{Department of Earth, Atmospheric and Planetary Sciences, Massachusetts Institute of Technology, Cambridge, MA 02139, USA}

\author[0000-0003-0525-9647]{Zifan Lin} 
\affiliation{Department of Physics, Washington University, St. Louis, MO 63130, USA} 
\affiliation{McDonnell Center for the Space Sciences, Washington University, St. Louis, MO 63130, USA}

\author[0000-0002-8119-3355]{Avi M. Mandell}
\affiliation{NASA Goddard Space Flight Center, Greenbelt, MD 20771, USA}

\author[0000-0003-0814-7923]{Elijah Mullens} 
\affiliation{Department of Astronomy and Carl Sagan Institute, Cornell University, 122 Sciences Drive, Ithaca, NY 14853, USA} 

\author[0000-0002-1046-025X]{Sarah Peacock}
\affiliation{NASA Goddard Space Flight Center, Greenbelt, MD 20771, USA}
\affiliation{University of Maryland, Baltimore County, Baltimore, MD, 21250, USA}

\author[0000-0002-2949-2163]{Edward W. Schwieterman} 
\affiliation{Department of Earth and Planetary Sciences, University of California, Riverside, CA, USA} 

\author[0000-0003-3305-6281]{Jeff A. Valenti}
\affiliation{Space Telescope Science Institute, 3700 San Martin Drive, Baltimore, MD 21218, USA} 

\author{{C. Matt Mountain}}
\affiliation{Association of Universities for Research in Astronomy, 1331 Pennsylvania Avenue NW Suite 1475, Washington, DC 20004, USA}

\author[0000-0002-3191-8151]{{Marshall Perrin}}
\affiliation{Space Telescope Science Institute, 3700 San Martin Drive, Baltimore, MD 21218, USA}

\author[0000-0001-7827-7825]{{Roeland P. van der Marel}}
\affiliation{Space Telescope Science Institute, 3700 San Martin Drive, Baltimore, MD 21218, USA}
\affiliation{William H. Miller III Department of Physics and Astronomy, Johns Hopkins University, Baltimore, MD 21218, USA}

\collaboration{100}{Consortium on Habitability and Atmospheres of M-dwarf Planets (CHAMPs)}


\begin{abstract}

Theoretical studies have suggested using planetary infrared excess (PIE) to detect and characterize the thermal emission of transiting \textit{and non-transiting} exoplanets, however the PIE technique requires empirical validation. Here we apply the PIE technique to a combination of JWST NIRSpec G395H transit and eclipse measurements of WASP-17b, a hot Jupiter orbiting an F-type star, obtained {consecutively (0.5 phase or 1.8 days apart)} as part of the JWST-TST program to perform Deep Reconnaissance of Exoplanet Atmospheres through Multi-instrument Spectroscopy (DREAMS). Using the in-eclipse measured stellar spectrum to circumvent the need for ultra-precise stellar models, we extract the first JWST nightside emission spectrum of WASP-17b using only transit and eclipse data thereby performing a controlled test of the PIE technique. 
From the WASP-17b nightside spectrum, we measure a nightside equilibrium temperature of $1005 \pm 256$ K and find tentative evidence for nightside \ce{SO2} absorption ($\ln B = 1.45$, $2.3\sigma$). 
In context with the dayside, the temperature of the nightside is consistent with (1) previous eclipse mapping findings that suggest relatively inefficient day-night heat transport, and (2) a non-zero bond albedo of $0.42^{+0.06}_{-0.10}$. 
\ce{SO2} on the nightside, if confirmed, would represent the first direct evidence for transport-induced chemistry, matching previous model predictions, and opening a new door into the 3D nature of giant exoplanets. 
{Our results suggest that PIE is feasible with JWST/NIRSpec for two epochs separated in time by significantly less than the rotation period of the host star.}

\end{abstract}


\section{Introduction}
\label{sec:intro}

Despite the prevalence and success of transiting exoplanet science, most exoplanets do not transit their host stars; particularly long period planets. Thus to fully explore the population of exoplanets in the solar neighborhood, methods must be developed and missions deployed to detect and characterize non-transiting exoplanets. The Planetary Infrared Excess \citep[PIE;][]{Stevenson2020} method presents an opportunity to study the atmospheres of non-transiting planets via combined light measurements.  Removing the large stellar contribution to the spectrum, which dominates the flux measurement at short wavelengths (e.g., visible and near-infrared), enables the extraction and study of the planetary emission spectrum at longer wavelengths (e.g., long-wavelength near-infrared and mid-infrared). Numerous theoretical works have demonstrated the potential for PIE using models and simulated data \citep{Stevenson2020, Lustig-Yaeger2021, Limbach2022, Mandell2022, Mayorga2023}. In particular, \citet{Lustig-Yaeger2021} showed that JWST should be capable of PIE studies of hot Jupiters; however, to date, this has not been demonstrated with empirical data. In this paper, we take the first step towards demonstrating and validating the fundamentals of PIE using JWST observations of the transiting exoplanet WASP-17b.  

The PIE technique critically hinges upon the ability to remove the stellar spectrum from combined-light measurements, so that only the spectrum of the planet remains. This is enabled by the fact that a cooler planet contributes negligibly to the combined system flux at sufficiently short wavelengths where a stellar model can be fitted to the data \citep{Stevenson2020}, thus allowing the planetary spectrum to emerge from the stellar model residuals at the longer wavelengths. The stellar models must, to first order, fit the stellar spectrum well enough for the planet flux to exceed the residuals on the fit. So while in theory JWST has the precision to measure PIE \citep{Lustig-Yaeger2021}, stellar models have not been demonstrated to this level of precision. Instead, in the era of JWST-quality NIR spectroscopy, insufficient complexity in stellar models is now a common occurrence \citep[e.g.,][]{Wakeford2019, Iyer2020, Garcia2022, Rackham2024}. Several efforts to improve stellar models are underway \citep[e.g.,][]{Witzke2022, Smitha2025}, notably to help confront the transit light source \citep[TLS;][]{Rackham2017, Rackham2023} effect that presently frustrates precise transmission spectroscopy observations around cool stars \citep{Moran2023, May2023, Lim2023, Cadieux2024, Canas2025, Fauchez2025, Radica2025}, but methods are needed to circumvent the issue of stellar models and to validate and leverage PIE observations with JWST. 

Fortunately, transiting planets offer an opportunity to validate core capabilities of the PIE technique without the use of stellar models. For transiting exoplanets, the true stellar spectrum can be precisely measured during secondary eclipse and then used as an empirical ``model'' for the star to validate and demonstrate the PIE technique without the need for ultra-precise and tunable theoretical stellar spectral models. This approach requires secondary eclipse observations at the same wavelengths and using the same instrument as observations at another orbital phase, such as primary transit when the planetary nightside is in view. Nightsides are compelling in and of themselves; studying exoplanet nightsides offers invaluable insights into 3D climate and chemistry, but historically have required time-intensive phase curves \citep[e.g.,][]{Stevenson2014c, Beatty2019, MikalEvans2023, Bell2024}. {Recent work has demonstrated that ground-based high-resolution spectroscopy is a viable and cost-effective alternative to characterize non-transiting planets \citep{Pelletier2021, Matthews2024} and the nightsides of transiting planets \citep{Yang2024, Mraz2024}. Additional strategies are needed to complement these efforts. } 
{We posit that given the improved spectroscopic capability, sensitivity, and instrument stability of JWST over past missions (e.g., Spitzer), along with the high oversubscription rate, it is worth revisiting multi-epoch phase measurements \citep[e.g.,][]{HarringtonEtal2006sciuandbphas, Cowan2007, Crossfield2010, Krick2016, Arcangeli2021} to examine if they may be a reliable means to efficiently obtain phase-dependent PIE emission spectra.}


WASP-17b presents an excellent opportunity to apply this non-model-dependent recipe for PIE to JWST data, not only to validate this flavor of PIE, but also to evaluate the use of PIE to study nightsides of hot Jupiters as well as to gain novel constraints on the WASP-17b nightside. 
WASP-17b is an inflated hot Jupiter (${\rm R_p} = 1.932 \, {\rm R_J}$) on a retrograde orbit around an F6 dwarf star (${\rm R_s} = 1.49 \, {\rm R}_{\odot}$) with an orbital period of 3.735 days \citep{Anderson2011, Alderson2022}. {The stellar rotation period of WASP-17 has not been confidently determined due to its low photometric variability \citep{Anderson2010}, although the best estimate from high-resolution spectroscopy places it near 8 days \citep{Anderson2011}.}  
Prior to JWST, observations with Hubble, Spitzer, and ground-based facilities indicated that WASP-17b possesses a highly extended atmosphere containing water vapor, Na and K \citep{Wood2011, Mandell2013, Wakeford2016, Sing2016, Sedaghati2016, Alderson2022}. Theoretical studies predict significant day-night and morning-evening variations in cloud coverage and atmospheric properties for WASP-17b \citep{Kataria2016, Zamyatina2023}. 

WASP-17b was observed during transit and eclipse with NIRISS SOSS \citep[for instrument details, see][]{Doyon2023}, NIRSpec G395H \citep[for instrument details, see][]{Jakobsen2022}, and MIRI LRS \citep[for instrument details, see][]{Kendrew2015} as part of the Deep Reconnaissance of Exoplanet Atmospheres using Multi-instrument Spectroscopy (DREAMS) program \citep[GTO-1353; PI Lewis;][]{Lewis_jwst.prop.1353L}. Several studies have already been conducted on these WASP-17b JWST measurements.   
\citet{Grant2023} found evidence of quartz clouds (\ce{SiO2(s)}) via an absorption feature at 8.6 \microns{} in the MIRI/LRS transmission spectrum. 
\citet{Louie2024} analyzed the NIRISS SOSS transmission spectrum and reported a supersolar water abundance and H$^{-}$ opacity. 
\citet{Gressier2025} investigated the NIRISS SOSS emission spectrum and found strong evidence of \ce{H2O} in the dayside atmosphere, indicative of a supersolar dayside \ce{H2O} abundance, as well as enhanced emission in the optical due to either a high internal temperature or reflected light. 
\citet{Valentine2024} studied the MIRI/LRS dayside eclipse data and found evidence for inefficient global heat redistribution that is consistent with a day–night temperature contrast of about 1000 K, and an eastward longitudinal hotspot offset of ${18.7}_{-3.8}^{+11.1\circ }$. 
Parallel ongoing work by \citet{lewis2025_w17transit} and \citet{wakeford2025_w17emission} report on the latest data analyses of the transit and eclipse measurements of WASP-17b with NIRSpec G395H.  
In sum, this set of observations from multiple instruments for the same hot Jupiter presents a valuable testbed with which to evaluate the PIE method using {the minimum set of multi-epoch phase measurements (just two epochs per instrument)}.

In this paper, we build on the recent JWST analyses of WASP-17 by reanalyzing the precise data, identifying the day and nightside PIE signals by dividing out the in-eclipse measurements, and interpreting the resulting planetary spectra using forward and retrieval models. 
The organization of this paper is as follows. Section \ref{sec:data} concerns the data and describes the JWST observations (\S \ref{sec:data:jwst}), data reduction (\S \ref{sec:data:reduction}), and in-eclipse spectrum removal method (\S \ref{sec:data:esr}) that is used to obtain WASP-17b's dayside and nightside PIE spectra (\S \ref{sec:data:pie}). Section \ref{sec:modeling} concerns atmospheric modeling, including atmospheric retrievals (\S \ref{sec:modeling:retrievals}) and photochemical models (\S \ref{sec:modeling:chemistry}). We discuss the implications of our findings in Section \ref{sec:discuss}, including the validity of the PIE technique (\S \ref{sec:discuss:PIE}), insights into the nightside (\S \ref{sec:discuss:nightside}) and caveats, challenges and prospects for future work (\S \ref{sec:discuss:caveats}).   




\section{Data} 
\label{sec:data}

Our treatment of the raw WASP-17b JWST data follows relatively standard data reduction approaches, as discussed below. However, we diverge from common practice with our parallel reductions of the transit and eclipse visits, which follow near-identical procedures to enable our in-eclipse spectrum removal approach (\autoref{sec:data:esr}) that is used to extract the planetary day and nightside emission spectra (\autoref{sec:data:pie}).   

\subsection{JWST Observations}
\label{sec:data:jwst}

WASP-17b was observed in transit with the G395H grating in the NIRSpec instrument \citep{Jakobsen2022} as part of the JWST GTO Cycle 1 Program 1353 (PI: N. Lewis; \citealp{Lewis_jwst.prop.1353L}) on 2023-03-31--14:45:54 UTC. Nearly two days later, on 2023-04-02--11:35:14 UTC, WASP-17b was observed in eclipse with the same instrument mode and setup. Both observations used the SUB2048 subarray with the NRSRAPID readout pattern and used 66 groups per integration for a total of  12.42 hours\footnote{The combined sets of transit and eclipse data for NIRSpec G395H used in this paper are available at \dataset[DOI: 10.17909/t5qk-ey95]{http://dx.doi.org/10.17909/t5qk-ey95}.}. Individual analyses of the WASP-17b transmission and emission spectra are provided in the companion papers \citet{lewis2025_w17transit} and \citet{wakeford2025_w17emission}, respectively. In this work we focus on the combined analysis that enables the extraction and study of the nightside spectrum of WASP-17b. To do this, we perform a set of standard data reduction steps for both the transit and eclipse datasets, making sure that both are subjected to the same procedures, as described next.    

\subsection{Data Reduction}
\label{sec:data:reduction}

We used the open-source, established {\eureka} \citep{Bell2022} pipeline to reduce the JWST time-series data, starting from the {\em uncal} FITS files and stepping through to generating planetary spectra (i.e., Stages 1-6). For this study of WASP-17b, we adopted similar methods to those used for previous transmission and emission analysis efforts (e.g., \citealt{ERSTeam2023, Ahrer2023, Alderson2023, FuLustig2023, Moran2023, Rustamkulov2023, Valentine2024}).

\subsubsection{Transmission and Emission Spectral Extraction}

For the transit and secondary eclipse data, we began with {\eureka}’s Stage 1 and the {\em uncal} FITS files, performing group-level background subtraction before determining the flux per integration using standard functions within the \texttt{jwst} pipeline. For the transit data, we masked a region within 8 pixels of the trace for pixel columns 600--2042 as is typical for {\eureka} NRS1 analyses, while for the secondary eclipse data, we masked within 8 pixels of the trace for all columns, per the standard {\eureka} NRS2 reduction process. In both cases, we estimated the per-column background as the median of the remaining pixels in a column with a 3$\sigma$ outlier rejection threshold and used a jump detection threshold of 15$\sigma$ in Stage 1.

We then processed the \texttt{jwst} Stage 1 {\em rateints} FITS output files through the regular \texttt{jwst} Stage 2 pipeline, skipping the background subtraction, flat fielding, and absolute photometric calibration steps when deriving the planet’s spectrum. In Stage 3 of {\eureka}, we turned off the background subtraction (\texttt{bg\_deg = None}) because we already subtracted the background at the group level in Stage 1. We used optimal spectral extraction \citep{Horne1986} with an aperture half-width of 5 pixels from the trace in all cases, which produced the lowest overall median absolute deviation (MAD) for the white light curve (WLC) reductions. When constructing the median frame for optimal spectral extraction, we applied an outlier rejection threshold of 15$\sigma$ for both datasets. During spectral extraction, we used a threshold of 10$\sigma$ for the transit data and 15$\sigma$ for the secondary eclipse data. All analyses used the Calibration Reference Data System (CRDS) \texttt{pmap} 1256, except the NRS2 eclipse, which used the 1258 \texttt{pmap} due to an overnight update. The differences between these two \texttt{pmap} values do not apply to the settings used for these analyses and therefore do not impact our results. We used \texttt{jwst} pipeline version 1.12.2 for these analyses.

\subsubsection{PIE Modifications}
\label{sec:data:reduction:piemod}

In addition to the standard \eureka reduction, a few important and specific data reduction steps were taken in Stages 1-3 to ensure consistency and validity in our subsequent analyses that synthesize multi-epoch measurements. Critically, Stages 1-3 should be consistent between transit and eclipse visit observations. The NRS1 and NRS2 analyses may have minor differences, but the transit and eclipse \eureka control files (ECFs\footnote{See https://eurekadocs.readthedocs.io/en/stable/ecf.html}) should look nearly identical, except for the files they point to. Particular care should be taken to ensure that the \texttt{spec\_hw} (aperture halfwidth), \texttt{xwindow} (horizontal subarray), and \texttt{pmap} (CRDS Context pmap) parameters are consistent between transit and eclipse, as these are the primary parameters that govern the signal vs background flux captured in the light curve reduction. By default, \eureka will use the latest \texttt{pmap} that is  available. This specifies the set of calibration files that should be used within the \texttt{jwst} pipeline.  However, if the transit and eclipse observations were not analyzed at the same time, then it is possible to get different calibration files. Therefore, explicitly specifying \texttt{pmap} in \eureka S1 is advised. 

\subsection{Calibrated Stellar Spectrum} 
\label{sec:data:cal} 

Extracting the calibrated stellar spectrum involves a few changes to the above description.  In Stage 2, we no longer skip the {\em photom} step, which applies flux calibrations to the data products and yields flux densities (in units of MJy).  In Stage 3, we set the keyword ``calibrated\_spectra'' to True and skip the additional background subtraction step.  We then run Stage 4cal of the {\eureka} pipeline to generate mean in- and out-of-eclipse stellar spectra.

\begin{figure}[t!]
    \centering
    \includegraphics[width=1.0\linewidth]{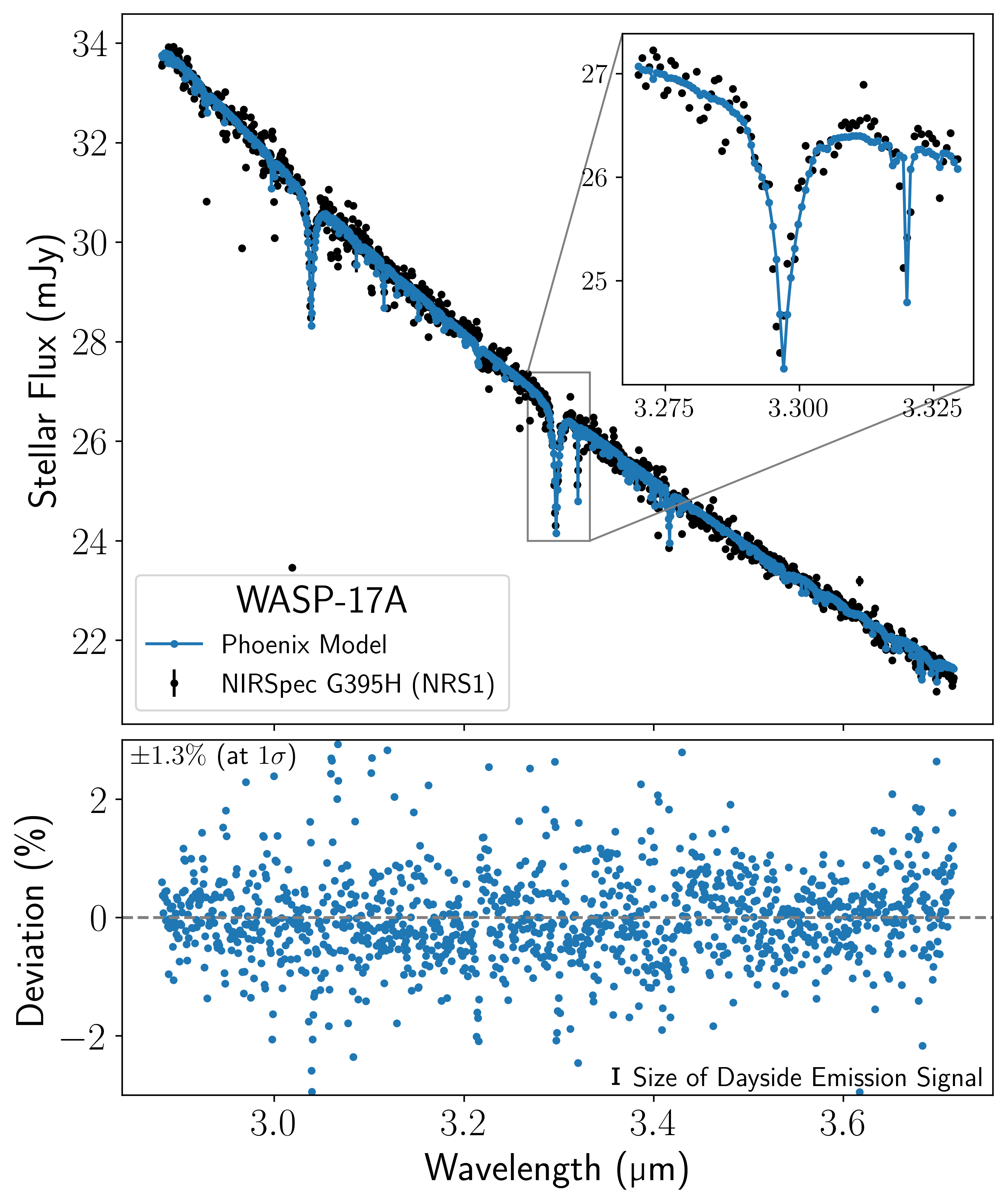}
    \caption{PHOENIX stellar model fit (blue points) to the calibrated stellar spectrum of WASP-17A observed with NIRSpec G395H using the NRS1 detector (black points). The relative residuals between data and model are scattered around zero with a standard deviation of ${\sim}1.3\%$. Thermal emission from the planet must exceed this data-model noise floor to extract the PIE signal. In this case, WASP-17b's average dayside planet-to-star flux ratio in NRS1 is measured at 0.133\% (indicated in lower panel), approximately $10\times$ below our model-derived limit.}
    \label{fig:calibrated_star}
\end{figure}

Fits to the observed stellar spectrum help to illuminate how the accuracy of stellar model grids is one of the primary challenges of the PIE technique at present. \autoref{fig:calibrated_star} shows the calibrated stellar spectrum of WASP-17A from the \eureka reduction (NRS1 only) compared with the best fitting PHOENIX stellar model \citep{pysynphot, STScI2018, Allard2012, Husser2013} that is based on known stellar parameters for WASP-17A ($R_s = 1.49~R_{\odot}$, $T_{\rm eff} = 6550$ K, $\log([ \rm M/H]) = -0.25$, $\log(g) = 4.2$; \citealp[]{Stassun2017}). Error bars on the calibrated stellar spectrum are derived from the variance in the time-series data, but are smaller than the size of the data points. Uncertainties from absolute flux calibration are not shown. The relative residuals between data and model are scattered around zero in the range of 1-2\% (with a standard deviation of ${\sim}1.3\%$). To first order, these residuals must be smaller than the thermal emission signal expected for the planet, at any given epoch, for the emission to be detectable. WASP-17b's average dayside planet-to-star flux ratio in NRS1 is measured at 0.133\%, approximately $10\times$ below our data-model residual noise limit. The nightside emission is expected to be an even smaller signal (and in the next section we show that it is). Thus, while the model provides an acceptable fit to the data, there is still considerable structure in the stellar spectrum (evident in all integrations) that is not well captured by the model and which ultimately limits our ability at present to remove the star with sufficient precision to capture the PIE signal. Subtle differences in the wavelength solution may also contribute to discrepancies in the fit.   Based on the above challenges, we pursue an alternative approach wherein the in-eclipse stellar spectrum is used as an empirical model to remove the stellar signal during other observational epochs when the planet is no longer hidden behind the star.    

\subsection{In-Eclipse Spectrum Removal (iESR)} 
\label{sec:data:esr} 

We begin with the \eureka Stage 3 output time-series spectra obtained during the separate eclipse and transit visits. Excluding ingress and egress, we break each time-series into three time periods: pre-eclipse, in-eclipse, and post-eclipse. Nominally, the pre- and post- combined light spectra should be the same with the exception of time-dependent effects, such as instrument systematics, stellar variability, and planetary rotation. Therefore, we elect to keep them separate throughout our initial analysis for diagnostic purposes. We then compute the weighted average spectrum during each time period for both transit and eclipse visits. Uncertainties are slightly inflated at this stage by taking the errors to be the quadrature sum of the coadded errors from averaging the time series spectra and the standard deviation of the fluxes in each spectroscopic channel. 

We now leverage the fact that the in-eclipse spectrum is only the spectrum of the star, $F_{\star}$, while the baseline spectra are combined light measurements of the planet and star, $F_{\rm baseline} = F_{\rm planet} + F_{\star}$. Under the assumption that the star remains stable over the course of an observation and, more critically, between separate visits, we use the following relations to extract planet-to-star flux ratios when different faces of the planet are visible to the observer,  
\begin{align}
    \frac{F_{\rm day1}}{F_{\star}} &= \frac{F_{\rm pre-eclipse}}{F_{\star}} - 1 \\ 
    \frac{F_{\rm day2}}{F_{\star}} &= \frac{F_{\rm post-eclipse}}{F_{\star}} - 1 \\ 
    \frac{F_{\rm night1}}{F_{\star}} &= \frac{F_{\rm pre-transit}}{F_{\star}} - 1 \\ 
    \frac{F_{\rm night2}}{F_{\star}} &= \frac{F_{\rm post-transit}}{F_{\star}} - 1.  
\end{align}
We note that if the stellar spectrum does change between visits, then the assumed nightside spectrum ${F_{\rm night}}/{F_{\star}}$ will be contaminated by the added term
\begin{equation}
\label{eqn:contamination}
    \epsilon = \frac{F_{\star,T}}{F_{\star,E}} - 1
\end{equation}
where $F_{\star,T}$ and $F_{\star,E}$ are the stellar spectra during the transit and eclipse visits, respectively. Thus, $\epsilon \rightarrow 0$, under our assumption that the star does not change between visits. For the F6 dwarf WASP-17A, this proves to be a good assumption. More generally $\epsilon$ can be used to estimate error tolerances relative to expected nightside signal sizes. We discuss the impact of nonzero $\epsilon$ values in our companion paper on HAT-P-26b \citep{hatp26b_pie}.  

Next we perform outlier rejection by sigma clipping spectral points at high resolution that deviate significantly from the average exhibited by the bulk of the data points. We attempted outlier rejection on the observed spectra prior to computing flux ratios, but found that a small number of significant outliers were created during the flux ratio step at wavelengths that were not outliers from the perspective of the individual visit spectra. Since more significant outliers exist in the nightside spectra (which rely on both transit and eclipse data) than the dayside spectra (which relies only on eclipse data), we attribute the outliers to instrument or stellar variations between visits. We tested multiple different metrics for outlier rejection, such as deviation from the mean of all spectral points (wavelength-independent) and deviation from the rolling median (highly chromatic). However, we opted to sigma clip high resolution spectral points that deviate significantly from the mean of points within lower resolution spectral bins. Using a low resolution spectral grid with $\Delta \lambda=0.02$ \microns{}, for each spectral point, we reject outliers at ${>} 3 \sigma$ from the mean in three iterative rounds of outlier rejection. This procedure removed 3\% of points from the dayside spectra and 4.5\% of points from the nightside spectra, with the majority of rejected points residing at the blue edge of the NRS1 detector. 

\subsection{Planetary Infrared Excess (PIE) from iESR} 
\label{sec:data:pie}

\begin{figure*}
    \centering
    \includegraphics[width = 0.99\textwidth]{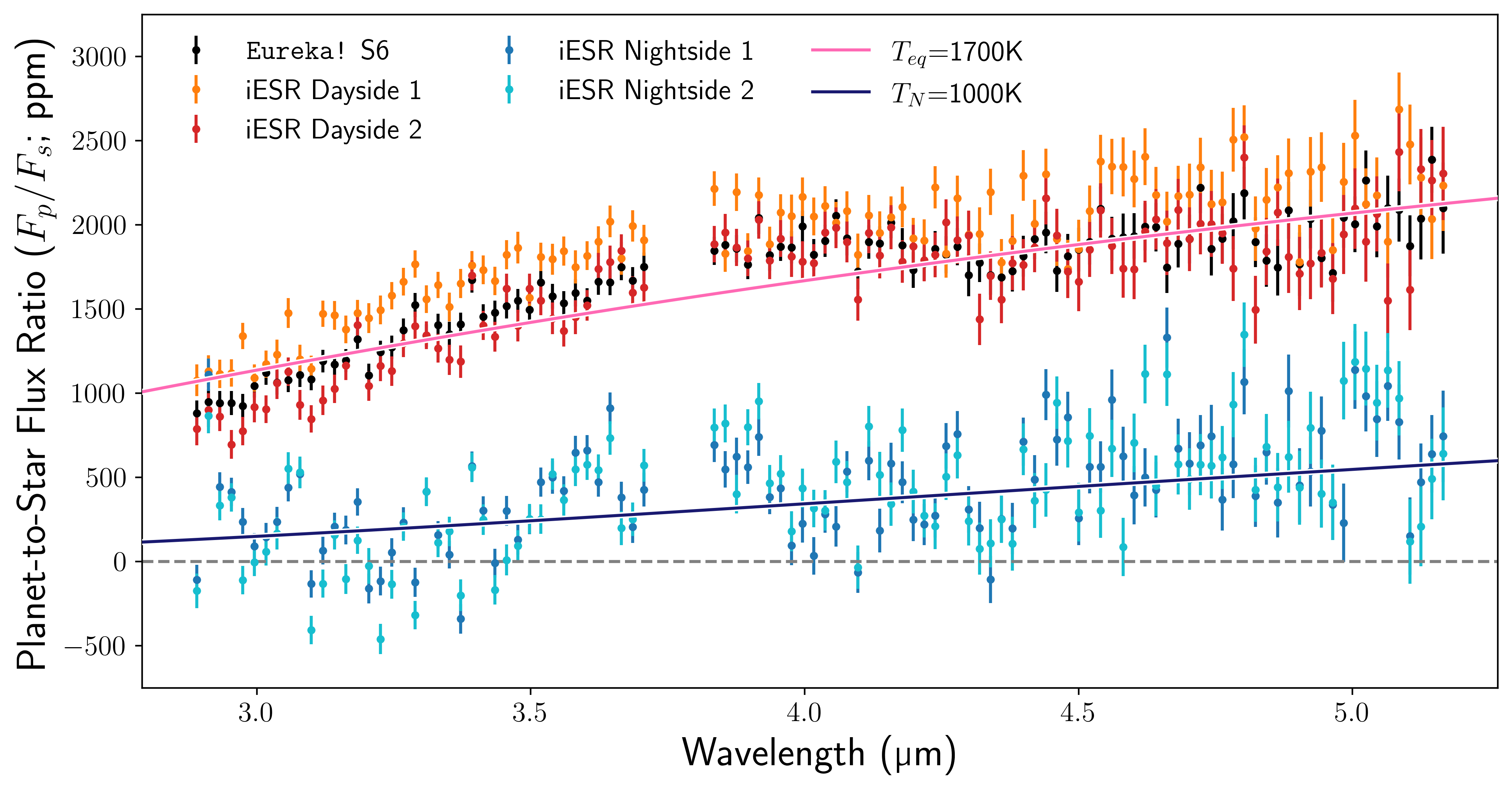}
    \caption{Planetary infrared excess spectra from the day (orange and red points) and nightsides (blue and cyan points) of WASP-17b. Dayside (nightside) 1 and 2 represent the pre-ingress and post-egress baseline spectra, respectively. The secondary eclipse dayside spectrum extracted using traditional methods is shown in black. Simple blackbody curves show the expected planet flux near the equilibrium temperature of WASP-17b (1700 K; pink line) and at 1000 K (navy blue line) for reference. }
    \label{fig:PIE1}
\end{figure*}


\autoref{fig:PIE1} shows the planet-to-star flux ratios extracted using iESR (see \autoref{sec:data:esr}) and reveals the planetary infrared excess from WASP-17b. The two dayside spectra are in excellent agreement with the dayside eclipse spectrum extracted from \eureka Stage 6 following traditional light curve fitting methods. One notable difference is that the pre-ingress (Dayside 1) spectrum exceeds the brightness of the post-egress (Dayside 2) spectrum, while the \eureka eclipse spectrum splits the difference. We measure a broadband brightness temperature in the G395H bandpass of $1814 \pm 45$ K for Dayside 1 and $1699 \pm 79$ K for Dayside 2, with a difference of $\Delta T = 115 \pm 91$ K. This modest difference is consistent with the offset hotspot for WASP-17b reported by \citet{Valentine2024} due to a super-rotating equatorial jet that pushes the planet's hot spot slightly east of the subsolar point. The shifted hotspot manifests in this emission spectrum offset because the hotspot is pointed more towards the observer, thereby contributing more to the disk-integrated hemispherical flux observed prior to the eclipse than after. We formally demonstrate the physical consistency of our finding using the MIRI eclipse map result from \citet{Valentine2024}. From MIRI, we calculate the median dayside temperature before and after eclipse, resulting in a pre eclipse brightness temperature of $1622 \pm 97$ K and a post eclipse brightness temperature of $1501 \pm 98$ K. While these temperatures are colder in MIRI than in NIRSpec due to the lower pressures probed, the pre-post eclipse temperature gradient ($\Delta T = 121$ K) is in excellent agreement with our observations in NIRSpec. This difference between pre-ingress and post-egress spectra 
{to infer a hot spot offset exemplifies the value of continuous partial phase curves in the JWST era \citep[see also,][]{Sikora2025}, building on pioneering work with HST \citep[e.g.,][]{KnutsonEtal2007natHD189733b}, \textit{Spitzer} \citep[e.g.,][]{Wong2014}.}
As expected, the average of the two iESR dayside spectra (not shown in \autoref{fig:PIE1}, but shown later in \autoref{fig:retrieval2}) is in excellent agreement with the standard \eureka eclipse spectrum, with differences $\le 1 \sigma$. This indicates that time-dependent instrumental or astrophysical effects are not compromising the dayside time series spectra, and physically meaningful spectra are being obtained.  

The nightside spectrum of WASP-17b is both novel and intriguing. The nightside spectra are much lower flux than the dayside, consistent with ${\sim}$1000 K blackbody emission. Unlike the dayside spectra, the Nightside 1 and 2 pre- and post-transit baseline spectra are in excellent agreement with one another, with no apparent brightness offsets between the two. We measure a broadband brightness temperature in the G395H bandpass of $1056 \pm 453$ K for Nightside 1 and $1040 \pm 591$ K for Nightside 2, an insignificant temperature difference albeit with greater uncertainties than the dayside measurements.  These nightside brightness temperature findings are consistent with expectations from general circulation models \citep{Kataria2016}. Perhaps most intriguingly, the WASP-17b nightside spectrum appears to contain several candidate absorption features not seen on the dayside. There is a coherent feature seen in both the dayside and nightside spectrum at 4.3 \microns{} where \ce{CO2} is well known to absorb \citep[e.g.,][]{ERSTeam2023, Rustamkulov2023}. We investigate these spectra with forward and inverse models in Section \ref{sec:modeling}.   

\subsection{Comparison between NIRSpec G395H, NIRISS SOSS, and MIRI LRS}
\label{sec:data:niriss_miri}


Thus far, we have focused exclusively on the NIRSpec G395H transit and eclipse data for WASP-17b. Although similar data exist from the NIRISS SOSS and MIRI LRS observations, initial investigations using the aforementioned iESR technique (\autoref{sec:data:esr}) were dominated by systematics that are common for these instruments. 

In particular, for the NIRISS SOSS measurements, differing wavelength solutions, 0th order contamination, and position angle changes between visits dominated over the faint emission signal expected from the nightside of WASP-17b at these shorter wavelengths. Figure 5 in \citet{Louie2024} shows the order 0 field star contaminant. We found that variations in the position of this contaminant relative to the spectral trace between the two visits generates a contaminating systematic effect for iESR PIE. {The NIRISS SOSS transit and eclipse observations were obtained nearly 32 days apart, which may have contributed to these challenges. If these or future observations were obtained closer in time, and hence with nearly identical position angles and contamination, then they might be suitable for PIE.}

While the expected emission from WASP-17b is greater at the longer wavelengths observed by MIRI LRS, long duration temporal ramps that differ between visits produces a differential effect that also dominates over the planetary signal. Figure 1 in \citet{Valentine2024} shows the wavelength dependence of these ramps in the MIRI LRS eclipse visit. By comparison, Figure 1 in \citet{Grant2023} shows how these ramps are different in the MIRI LRS transit visit. Ramp fitting can flatten the baseline trends, but doing so introduces uncertainty in the baseline flux zero point. Novel detrending methods, such those presented in \citet{Fortune2025} to down-weight specific pixels that exhibit increased systematics, may be key to enabling iESR for PIE at MIRI wavelengths.  

Based on these challenges using NIRISS SOSS and MIRI LRS, we elect to focus this paper on the comparatively more stable and reproducible NIRSpec G395H data for our WASP-17b nightside analysis, and we leave more detailed analyses of the systematic effects to future work. We discuss the caveats and implications of instrument systematics for future PIE studies in \autoref{sec:discuss:caveats}. 

\section{Modeling} 
\label{sec:modeling}

We perform a two-step atmospheric modeling procedure involving the use of atmospheric retrievals and chemistry modeling. First, retrievals are used to infer the nightside pressure-temperature (PT) profile and molecular abundance constraints from the PIE spectrum. Second, a grid of photochemical transport models are used to interpret the retrieved nightside atmospheric constraints to understand their physical and chemical relationship to the dayside findings.


\subsection{Atmospheric Retrieval Modeling}
\label{sec:modeling:retrievals}  

\subsubsection{Retrieval Setup}

We use the \poseidon\footnote{\texttt{https://github.com/MartianColonist/POSEIDON/tree/main/POSEIDON}} (v1.2) atmospheric retrieval model to explore a broad range of possible atmospheric model fits to the WASP-17b day and nightside emission spectra and perform Bayesian inference \citep{MacDonald2017, MacDonald2023, Mullens2024}. \poseidon was originally developed for exoplanet transmission spectroscopy retrievals, but has recently been extended to include basic emission spectroscopy capabilities \citep{Coulombe2023}. We configure \poseidon to use 1-D single-stream radiative transfer without multiple scattering. For more details about the radiative transfer prescription, we refer the reader to \citet{Coulombe2023}. We use opacity sampling with \poseidon to perform the radiative transfer calculations on an intermediate resolution spectral grid set to $R=60,000$ from 2.8 - 5.3 \microns{} throughout this work. We use the four-parameter double-gray ``Guillot'' temperature-pressure profile parameterization adapted from \citet{Guillot2010} \citep[see also, ][]{Molliere2019, Mullens2024}. Opacities are used for the following species (from \poseidon v1.2): \ce{H2O} \citep{Polyansky2018}, \ce{CH4} \citep{Yurchenko2024}, \ce{CO2} \citep{Yurchenko2020}, \ce{CO} \citep{Li2015}, \ce{SO2} \citep{Underwood2016}, \ce{H2S} \citep{Azzam2016}, and \ce{NH3} \citep{Coles2019} as well as collision-induced absorption (CIA) from HITRAN \citep{Karman2019} for \ce{H2-H2}, \ce{H2-He}, \ce{H2-CH4}, \ce{CO2-H2}, \ce{CO2-CO2}, and \ce{CO2-CH4}. The volume mixing ratio (VMR) of each molecule is included as a free parameter and assumed to be evenly-mixed throughout the atmospheric vertical column. 

\poseidon uses the nested sampling Bayesian parameter estimation model \texttt{MultiNest} \citep{Feroz2009}, which is implemented via the \texttt{PyMultiNest} Python wrapper \citep{Buchner2014}. Our retrievals use 400 live points and contain 12 free physical parameters and, as we discuss below, an additional free parameter for error inflation. A summary of the free parameters used in our retrievals and their respective priors are listed in the second column of Table \ref{tab:RetrievalResults}.  

We run retrievals with and without an error inflation term included to account for additional sources of uncertainty that may be unquantified within the data. We use the approach of \citet{Line2015} (included within \poseidon), which uses the following effective error in the retrieval 
\begin{equation}
\label{eqn:b_error}
    {s}_{i}^{2}={\sigma }_{i}^{2}+{10}^{b},
\end{equation} 
where the photometric errors and the error inflation (squared) term, $\sigma_{\rm infl}^2=10^{b}$, are summed in quadrature for the $i^{\rm th}$ spectral bin. \poseidon fits directly for $b$, modifying the data uncertainties in real time while it fits the spectrum and infers physical parameters. We use a broad uniform prior on $b$ bounded on the low end three orders of magnitude smaller than the smallest error bar in the spectrum and on the high end two orders of magnitude larger than the largest error bar. That is,      
\begin{equation} 
\label{eqn:prior_b}
    \mathcal{P}(b) \sim \mathcal{U}(\log_{10}[\min(\sigma^2)] - 3, \log_{10}[\max(\sigma^2)] + 2). 
\end{equation} 
Finally, we retrieve on a single nightside (dayside) emission spectrum produced by averaging the pre- and post-transit (eclipse) baseline spectra shown in \autoref{fig:PIE1}. Both day and nightside spectra contain 105 data points.   

\subsubsection{Retrieval Results}

\begin{figure*}[t]
    \centering
    \includegraphics[width = 0.99\textwidth]{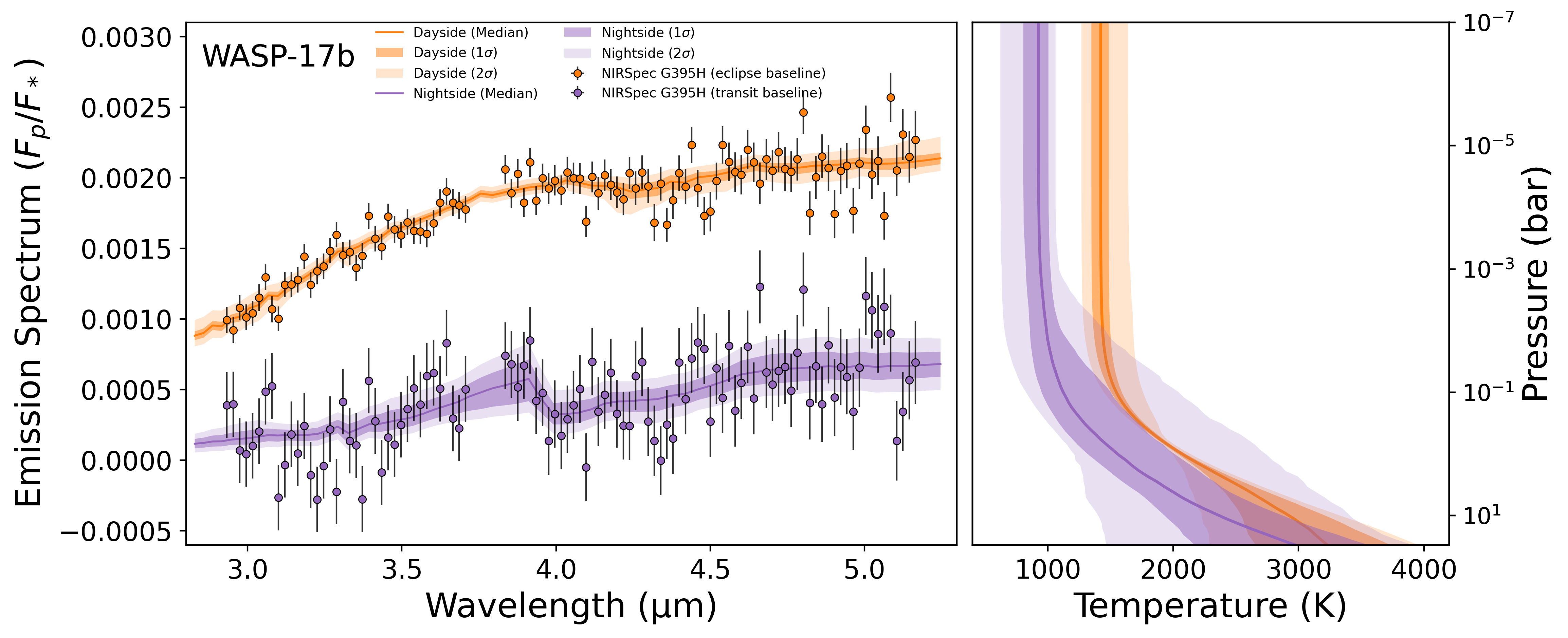}
    \includegraphics[width = 0.99\textwidth]{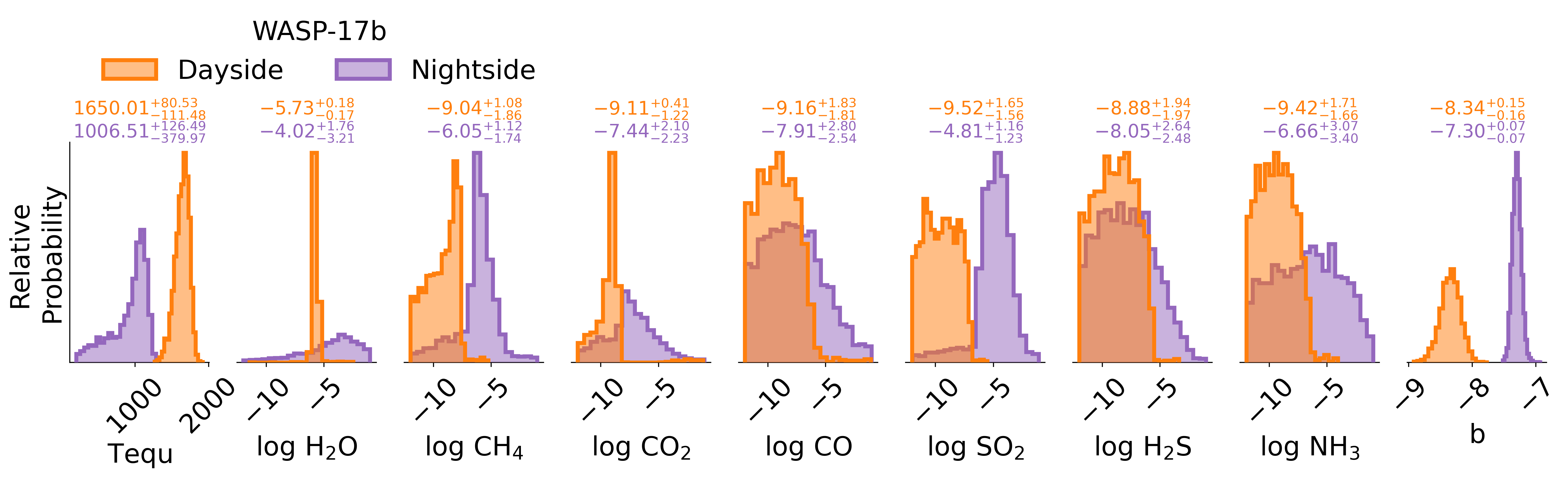}
    \caption{Summary of retrieval results for the day (orange) and night (purple) side emission spectra of WASP-17b for retrievals using error inflation. \textbf{Top Left:} Emission spectra measurements fitted with the median retrieved model spectrum with envelopes for the 1 and 2 sigma posterior credible intervals. Error bars shown are inflated using \autoref{eqn:b_error} with the median retrieved value of $b$.  \textbf{Top Right:} Retrieved PT profile constraints. \textbf{Bottom Row}: Histograms showing the one-dimensional posterior probability density for a subset of retrieved parameters. The subplot titles list the 1 sigma posterior constraints.} 
    \label{fig:retrieval2}
\end{figure*}

\autoref{fig:retrieval2} and \autoref{fig:retrieval2_no_infl} summarize our \poseidon retrieval results with the error inflation term and without it, respectively. \autoref{tab:RetrievalResults} also tabulates the $1{\sigma}$ posterior constraints on each parameter, the best-fit reduced chi-squared ($\chi^2_{\nu}$) statistic, and the log-evidence derived for the day and nightside retrievals.  

\begin{deluxetable*}{r|l||l|l|l|l}
\tablewidth{0.97\textwidth}
\tablecaption{\poseidon Retrieval Model Free Parameters \& $1\sigma$ Posterior Constraints  \label{tab:RetrievalResults}} 
\tablehead{
\colhead{} & \colhead{} & \multicolumn{4}{c}{Posteriors} \\
\cline{3-6} 
\colhead{Parameters} & \colhead{Priors} & \colhead{Dayside} & \colhead{Nightside} & \colhead{Dayside (Err. Infl.)} & \colhead{Nightside (Err. Infl.)}
}
\startdata
$\mathrm{R}_{\mathrm{p, \, ref}}$ & $\mathcal{N}(1.70, 0.02)$   & $1.699^{+0.017}_{-0.018}$ &   $1.703^{+0.018}_{-0.019}$ & $1.699^{+0.017}_{-0.018}$ & $1.701^{+0.018}_{-0.017}$ \\
$\log \, \kappa\mathrm{IR}$ & $\mathcal{U}(-5.00, 0.00)$        &   $-4.75^{+0.18}_{-0.14}$ &     $-3.42^{+0.40}_{-0.40}$ & $-4.80^{+0.17}_{-0.12}$ & $-4.31^{+1.05}_{-0.44}$ \\
$\log \, \gamma$ & $\mathcal{U}(-4.00, 1.00)$                   &   $-1.37^{+0.34}_{-0.39}$ &        $-1.8^{+2.1}_{-1.7}$ & $-1.27^{+0.36}_{-0.43}$ & $-1.52^{+0.92}_{-1.06}$ \\
$\mathrm{T_{\rm int}}$ & $\mathcal{U}(50.00, 2300.00)$                 &       $563^{+388}_{-332}$ &         $489^{+138}_{-117}$ & $621^{+406}_{-354}$ & $538^{+469}_{-330}$ \\
$\mathrm{T_{equ}}$ & $\mathcal{U}(200.00, 2300.00)$             &       $1617^{+88}_{-111}$ &         $366^{+168}_{-112}$ & $1649^{+80}_{-112}$ & $1005^{+129}_{-383}$ \\
$\log \, \mathrm{H_2 O}$ & $\mathcal{U}(-12.00, -1.00)$         &   $-5.78^{+0.13}_{-0.14}$ &        $-7.1^{+1.4}_{-3.1}$ & $-5.73^{+0.18}_{-0.17}$ & $-4.1^{+1.8}_{-3.2}$ \\
$\log \, \mathrm{CH_4}$ & $\mathcal{U}(-12.00, -1.00)$          &      $-8.1^{+0.3}_{-1.8}$ &  $-6.159^{+0.098}_{-0.092}$ & $-9.0^{+1.1}_{-1.8}$ & $-6.1^{+1.1}_{-1.8}$ \\
$\log \, \mathrm{CO_2}$ & $\mathcal{U}(-12.00, -1.00)$          &   $-9.01^{+0.27}_{-0.56}$ &     $-7.64^{+0.16}_{-0.15}$ & $-9.11^{+0.40}_{-1.24}$ & $-7.4^{+2.1}_{-2.2}$ \\
$\log \, \mathrm{CO}$ & $\mathcal{U}(-12.00, -1.00)$            &      $-9.4^{+1.8}_{-1.7}$ &     $-6.15^{+0.40}_{-0.56}$ & $-9.2^{+1.8}_{-1.8}$ & $-7.9^{+2.8}_{-2.5}$ \\
$\log \, \mathrm{SO_2}$ & $\mathcal{U}(-12.00, -1.00)$          &      $-9.5^{+1.6}_{-1.6}$ &     $-5.96^{+0.11}_{-0.11}$ & $-9.5^{+1.7}_{-1.6}$ & $-4.8^{+1.2}_{-1.2}$ \\
$\log \, \mathrm{H_2 S}$ & $\mathcal{U}(-12.00, -1.00)$         &      $-8.4^{+1.9}_{-2.3}$ &        $-9.3^{+2.0}_{-1.8}$ & $-8.9^{+2.0}_{-2.0}$ & $-8.0^{+2.6}_{-2.5}$ \\
$\log \, \mathrm{NH_3}$ & $\mathcal{U}(-12.00, -1.00)$          &      $-9.4^{+1.6}_{-1.7}$ &     $-5.78^{+0.29}_{-0.32}$ & $-9.4^{+1.7}_{-1.7}$ & $-6.7^{+3.0}_{-3.4}$ \\
$\mathrm{b}$ & $\mathcal{U}(-11.5, -5.4)$                        & ---                       &   ---                       & $-8.34^{+0.15}_{-0.16}$ & $-7.302^{+0.070}_{-0.069}$ \\
\hline
$\chi^2_{\nu}$ & --- & 1.877 & 8.330  & 1.288 & 1.220  \\
$\ln Z$ & --- & 772.97 & 469.44  & 778.76 & 713.61  \\
\enddata
\tablecomments{$\mathcal{U}$ and $\mathcal{N}$ denote uniform and normal prior distributions, respectively. The error inflation prior for the day and nightsides is derived from \autoref{eqn:prior_b}. 
Preferred nightside results use error inflation. For more on the preferred dayside results see \autoref{tab:RetrievalResultsDayside} in \autoref{appendix:more_retrievals}.} 
\end{deluxetable*} 

The inferred error inflation parameters are different between the day and nightside retrievals, which indicates that additional sources of error are present, and underscores the importance of including the error inflation term. Due to the high $\chi^2_{\nu}$ of the nightside retrieval when error inflation is not used (see \autoref{tab:RetrievalResults} and \autoref{fig:retrieval2_no_infl}), a higher value of $\sigma_{\rm infl} = 224 \pm 18$ ppm for the error inflation is required for the nightside spectrum compared to the derived value of $\sigma_{\rm infl} = 68 \pm 12$ ppm for the dayside. These $\sigma_{\rm infl}$ values (in ppm) represent an effective noise floor that adds in quadrature with the original error bars via \autoref{eqn:b_error}. 

It is not surprising that the nightside spectrum incurs larger errors than the dayside iESR PIE data as it requires data from two independent visits to be synthesized, as opposed to just the eclipse data for the dayside iESR analysis. While 224 ppm is a useful rule-of-thumb noise floor for the iESR PIE investigation that we present here, we caution that because it encodes errors associated with stellar variability and instrument systematic differences between visits, it is not a general result that can be directly applied to other systems. We revisit this point in the Discussion (\autoref{sec:discuss:caveats}). Ultimately, we consider the inclusion of the error inflation parameter to be critical for this analysis to account for the additional sources of error that cannot be readily quantified. Therefore, throughout the rest of the paper, we primarily focus on results that include error inflation, although we make comparisons to the retrievals without inflation where appropriate, and we present the full results in \autoref{appendix:more_retrievals}. 

\autoref{fig:retrieval2} and \autoref{tab:RetrievalResults} show that fits to the day and nightside emission spectra of WASP-17b reveal a cooler nightside with hints of different dominant molecular absorbers compared to the dayside. The dayside emission spectrum corresponds to an equilibrium temperature of $1650 \pm 95$ K while that of the cooler nightside is $1007 \pm 250$ K. While \ce{H2O} and \ce{CO2} are constrained on the dayside, on the nightside \ce{SO2} and \ce{CH4} are most tightly constrained and contribute to the absorption features seen in the spectrum.  


In addition to the posterior constraints in \autoref{tab:RetrievalResults}, we compute detection significances for individual molecules by running retrievals without each gas and calculating Bayes Factors relative to the baseline case with all of the gases. In the dayside NIRSpec G395H spectrum, without error inflation, we detect \ce{H2O} at $2.7\sigma$ ($\ln B = 2.26$), \ce{CO2} at $1.9\sigma$ ($\ln B = 0.77$), \ce{CH4} at $1.4\sigma$ ($\ln B = 0.21$). A combination of all gases is detected at $8.6\sigma$ ($\ln B = 34.94$). With the error inflation, we obtain similar results except 
the detection of any features from a combination of all free parameter gases falls to $4.7\sigma$ ($\ln B = 9.25$). 

We note that our dayside retrieval reveals the subtle presence of a second alternative posterior mode with a larger abundance of trace gases and a more isothermal thermal structure with two radiative zones. This type of degeneracy is not uncommon in emission retrievals with limited wavelength range and precision \citep[e.g.,][]{Waldmann2015, Gandhi2018, Mettler2024}. {The presence of two posterior modes contributes to the low significance individual gas detections reported above for the dayside spectrum, because the removal of individual gases can be roughly compensated for by modifying the PT profile structure.} To further investigate these two solutions we ran two additional retrievals with modified priors on $\log \gamma$ (i.e., the ratio between the visual and infrared opacity) that effectively bifurcate the two posterior modes into the following subcases: ``low $\gamma$'' where $-4.0 < \log \gamma < -0.5$ and ``high $\gamma$'' where $-0.5 < \log \gamma < 1.0$. The ``low $\gamma$'' case agrees with the fiducial model with a broad $\gamma$ prior while the ``high $\gamma$'' case has a lower intrinsic temperature, has a lower pressure photosphere, and allows higher trace gas abundances. However, despite the retrieval model's preference for the ``low $\gamma$'' case when using a broad prior on $\gamma$, parallel work studying the panchromatic WASP-17b dayside emission spectrum using NIRISS SOSS, NIRSpec G395H, and MIRI LRS data to cover wavelengths from $0.6 - 12$ \microns favor a PT profile that is consistent with our ``high $\gamma$'' subcase \citep{wakeford2025_w17emission}. This suggests that our ``high $\gamma$'' case may be more physically representative of the WASP-17b thermal structure and highlights the challenge of interpreting spectra in the G395 bandpass in isolation. Therefore, in our subsequent assessment of photochemistry transport modeling presented in \autoref{sec:modeling:chemistry}, we use the panchromatic dayside retrieval results from \citet{wakeford2025_w17emission} as the state of the art interpretation for the dayside chemistry of WASP-17b.  

In the nightside spectrum, whether or not we include error inflation plays a larger role than seen for the dayside. Without error inflation, we detect \ce{CH4} at $8.8\sigma$ ($\ln B = 36.49$), \ce{SO2} at $7.6\sigma$ ($\ln B = 26.83$), \ce{CO2} at $7.0\sigma$ and ($\ln B = 22.44$). A combination of all gases is detected at $14\sigma$ ($\ln B = 99.21$). With error inflation, these numbers drop significantly and we detect \ce{CH4} at $1.5\sigma$ ($\ln B = 0.28$), \ce{SO2} at $2.3\sigma$ ($\ln B = 1.45$), \ce{CO2} at $0.9\sigma$ ($\ln B = -0.12$), \ce{H2O} at $1.5\sigma$ ($\ln B = 0.34$), and a combination of all gases (i.e., features in the emission spectrum) are detected at $3.0\sigma$ ($\ln B = 3.23$). This sensitivity of the nightside gas detection significances to error inflation is not unexpected due to the much larger error inflation term inferred from these fits compared to the dayside spectrum, as well as the subsequent drop in $\chi^2_{\nu}$ from 8.3 to 1.2 and increase in $\ln Z$ from 469.4 to 713.6. These quantitative improvements inform our decision to favor the results with error inflation. We also note that the nightside retrievals showed no evidence for multi-mode solutions regardless of error inflation and across dozens of initial retrieval tests using different PT parameterizations and priors.     

\subsubsection{Pressure Contribution Functions}

The retrieval results illuminate the range of atmospheric pressures on the day and nightsides where the flux emerges, which can offer important context for understanding how the retrieved abundances compare to expectations from atmospheric forward models. We post-process our \poseidon retrieval results to generate pressure contribution contours for each gaseous species as a function of pressure and wavelength. This is done using \poseidon v1.2 with updates for pressure contribution functions described in \citet{Mullens2024}. We repeat this calculation for 100 equally weighted random samples from the posterior distribution to determine the average pressure contribution function that combine both the uncertainty in the atmospheric state (due to measurement uncertainty) and the range in emitting pressures that naturally contribute to the chromatic planet flux (due to radiative transfer in a non-isothermal gas). 

\begin{figure*}[t]
    \centering
    \includegraphics[width = 0.99\textwidth]{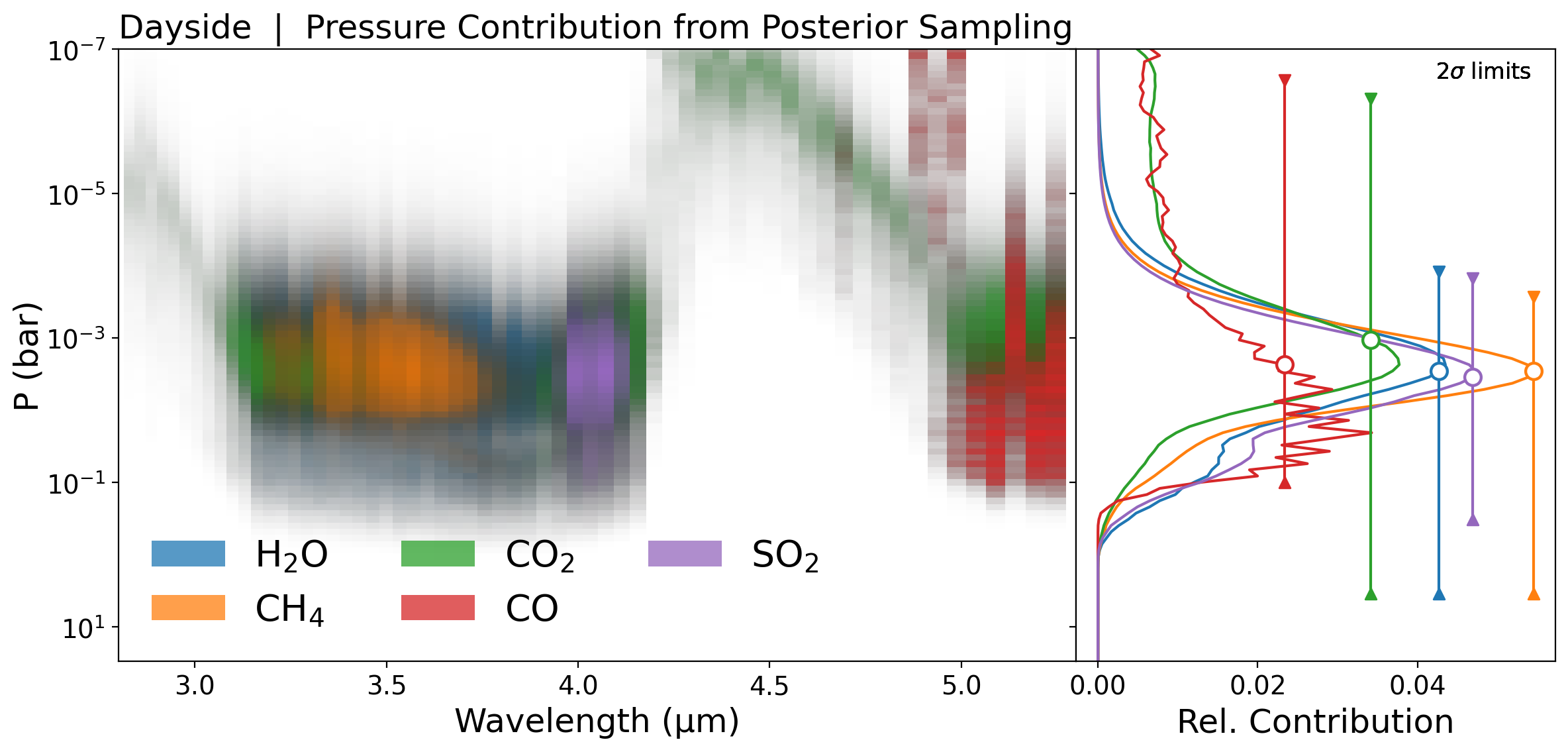}
    \includegraphics[width = 0.99\textwidth]{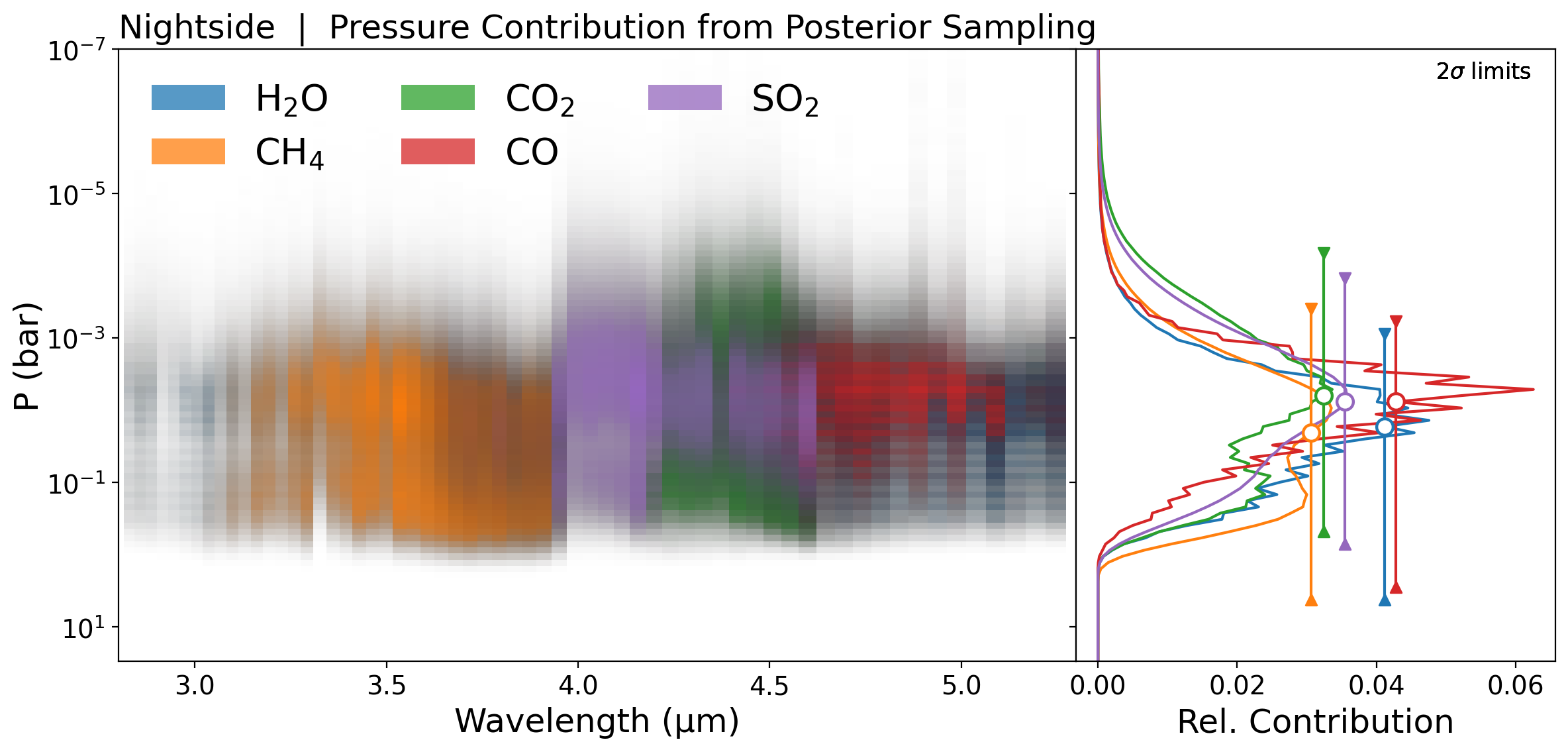}
    \caption{Pressure contribution functions from the retrieved day (top) and night (bottom) side emission spectra for the primary gaseous species in the retrieval models (colors). The left panels show pressure contributions as a function of wavelength while the right panels show the same contributions integrated over the NIRSpec G395H bandpass. Vertical error bars show the range of pressures where 95.5\% ($2 \sigma$) of the flux that is sensitive to each gas probes.} 
    \label{fig:pressure_contribution}
\end{figure*}

\autoref{fig:pressure_contribution} shows our resulting pressure contribution functions for the day (top panels) and night (bottom panels) sides of WASP-17b. As stated above, the range in pressures probed reflects both an intrinsic range based on the atmospheric thermal structure, abundances, and opacities, and the \textit{uncertainty} in our knowledge of these touchstone atmospheric characteristics. In general, the median day and nightside flux in the G395H bandpass appears to emanate from near 1 mbar with a range of 3-7 orders of magnitude in pressure to capture 95\% of the emission.

\subsection{Atmospheric Chemistry \& Transport Modeling}
\label{sec:modeling:chemistry}

\subsubsection{Two-Column Model Setup}

To compare the atmospheric compositions retrieved using the spectra from the iESR PIE technique with those predicted by physical models, we present a two-column photochemical model to explore plausible abundances on the dayside and nightside, regulated by global transport. This two-column photochemical model is adapted from the 2D VULCAN Photochemical Model \citep{Tsai2024}, simplified to represent only two vertical columns that correspond to the dayside and nightside hemispheres. We adopt the retrieved median temperature profiles for each hemisphere. With the same day-night temperature structure unchanged, we explore a range of metallicities (0.1, 1, 10, 50 $\times$solar), C/O ratios (0.1, 0.25, 0.549, 0.75), vertical mixing strengths (eddy diffusion coefficient $K_{{\rm zz}}$ = 10$^{7}$, 10$^{8}$, 10$^{9}$, 10$^{10}$, 10$^{11}$ cm$^2$/s), and zonal wind speeds (100, 200, 500, 1000, 3000, 5000 m/s). For simplicity, the eddy diffusion coefficient is taken to be constant throughout the atmosphere, and the zonal wind is assumed to be uniform above the 1-bar level (P $<$ 1 bar) and exponentially decaying to zero at higher pressures, consistent with our understanding from general circulation models \citep{Showman2020}. For the stellar UV flux of WASP-17, we use a semi-empirical F-star spectrum with $T_{\mathrm{eff}} = 6500$ K from \cite{Rugheimer2013}. A zenith angle of 58$^{\circ}$ is assumed for the dayside column, following \cite{Tsai2021}. 

\subsubsection{Metallicity constraints}

\begin{figure*}[t]
    \centering
    \includegraphics[width = 0.45\textwidth,trim={0.3cm 0cm 3.7cm 0cm},clip]{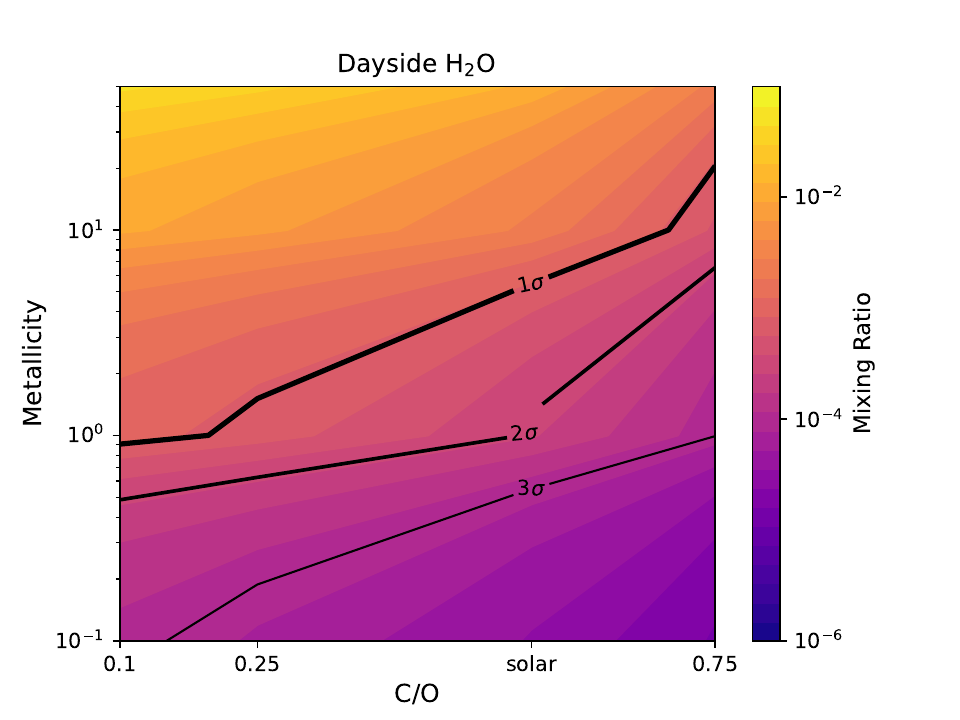}
    \includegraphics[width = 0.53\textwidth,trim={0.8cm 0cm 1cm 0cm},clip]{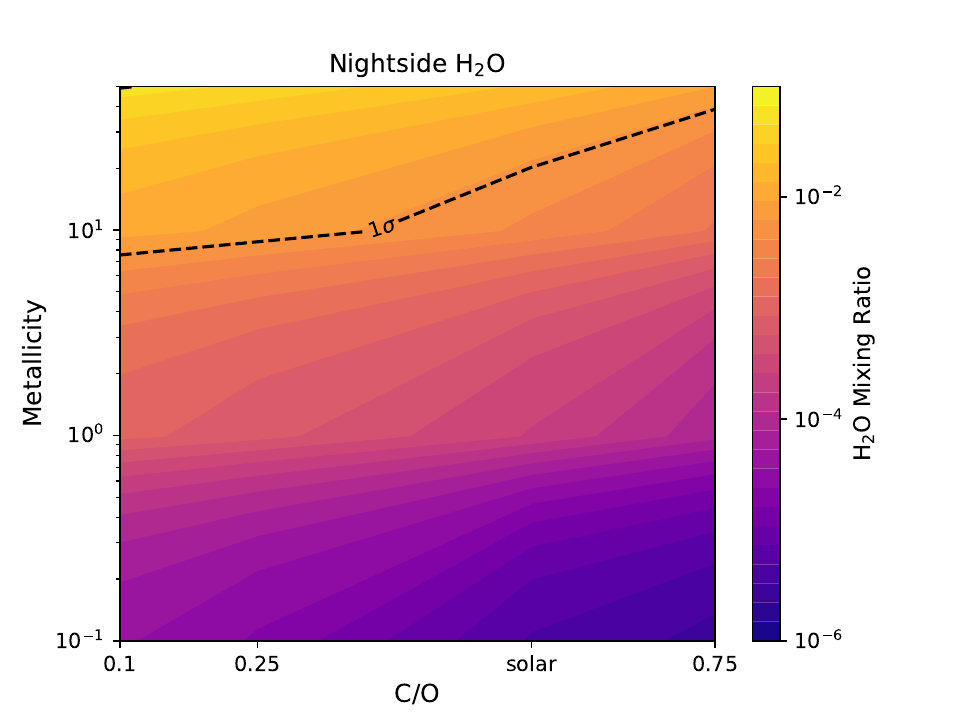}
    \caption{The dayside and nightside abundances of \ce{H2O} as a function of metallicity and C/O ratio computed from our two-column VULCAN grids{, assuming $K_{{\rm zz}}$ = 10 cm$^2$s$^{-1}$ and no day-night transport}. The abundances are averaged between 0.1 bar and 10$^{-4}$ bar,  guided by the contribution function shown in \autoref{fig:pressure_contribution}. {}
    The dashed lines and solid lines show the upper and lower bounds of the 1-, 2-, 3-$\sigma$ uncertainty intervals, respectively. Bounds outside of modeled abundance ranges are not shown (e.g., only the 1-$\sigma$ upper limit of \ce{H2O} is visible for the nightside).
    }
    \label{fig:h2o_contour_met_co}
\end{figure*}

{Despite low significance detections of molecules in our free-chemistry retrievals, these results still provide informative constraints on the volume mixing ratios of putative atmospheric constituents, such as} \ce{H2O}, \ce{CH4}, \ce{CO2}, and \ce{SO2}. Among them, the \ce{H2O} abundance is well constrained on the dayside and weakly constrained on the nightside. Since \ce{H2O} is also the dominant oxygen-bearing molecule and not expected to exhibit significant day-night variations \citep{Tsai2024,Zamyatina2023} {and insensitive to transport processes}, we first use \ce{H2O} to constrain atmospheric metallicity. \autoref{fig:h2o_contour_met_co} compares the retrieved \ce{H2O} abundances with model predictions across a range of metallicities and C/O ratios. Based on the 1$\sigma$ constraints, the atmospheric metallicity consistent with both dayside and nightside water abundances lies approximately between solar and 50$\times$ solar.


\subsubsection{Nightside \ce{SO2} requires horizontal transport}

We next interpret the tentative detection of \ce{SO2}. Our retrievals yield higher \ce{SO2} and tighter constraints on the nightside compared to the dayside. In hydrogen-dominated atmospheres, \ce{SO2} is primarily produced by photochemical processes \citep{Tsai2023a}. In the absence of interaction with the dayside, nightside sulfur is expected to remain predominantly in \ce{H2S}, while \ce{SO2} should be negligible. However, 2D photochemical modeling has predicted that the horizontal transport timescale is much shorter than the lifetime of \ce{SO2}, allowing it to accumulate on the nightside, even potentially reaching higher abundances than on the dayside due to lack of photodissociation \citep{Tsai2023c}.     

\begin{figure}[t]
    \centering
    \includegraphics[width = 0.49\textwidth]{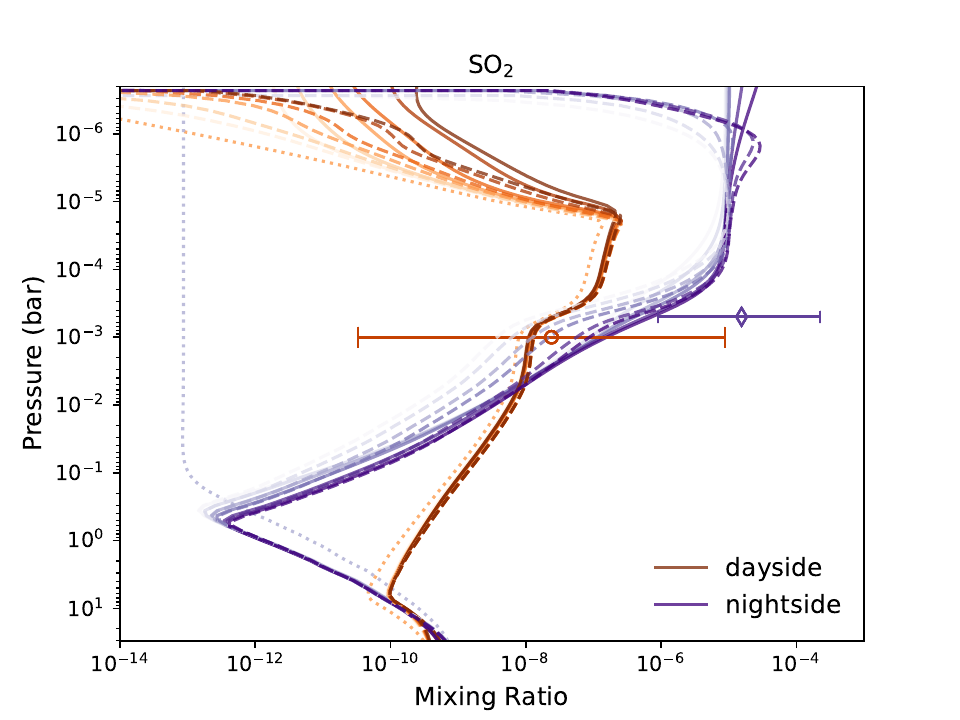}
    \caption{The dayside (orange) and nightside (purple) abundance profiles of \ce{SO2} from our two-column VULCAN grids with 10$\times$ solar metallicity. The solid lines assume $K_{{\rm zz}}$ = 10$^{11}$ cm$^2$/s, whereas the dashed lines assume $K_{\mbox{zz}}$ = 10$^7$ cm$^2$/s, showing the strongest and weakest vertical mixing cases in our model grid, respectively. The light to dark shades of each color represent zonal wind speed of 100, 200, 500, 1000, 3000, 5000 m/s. The dotted lines are those without including zonal wind (vertical mixing only). The error bars show 1-$\sigma$ retrieved constraints, with their vertical positions chosen to be near the peak of the contribution function.}
    \label{fig:so2_wind}
\end{figure}

\autoref{fig:so2_wind} illustrates that the dayside and nightside \ce{SO2} abundances informed by our retrievals are consistent with a broad range of horizontal wind speeds. Since the lifetime of \ce{SO2} ($\gtrsim$10 days) tends to be longer than the transport timescale, the resulting \ce{SO2} profiles are insensitive to the zonal wind speed. However, without any zonal wind, the nightside \ce{SO2} (purple dotted line in \autoref{fig:so2_wind}) is negligible, as expected, and strongly disagrees with the retrieved abundance of ${\sim} 15$ ppm. This evidence of non-negligible zonal wind is consistent with the hot spot inferences of \citet{Valentine2024}. The modeled \ce{SO2} abundances are not too sensitive to vertical mixing across a wide range of $K_{{\rm zz}}$ (solid vs dashed lines). This emphasizes the critical role of zonal winds in understanding nightside \ce{SO2}. Within the explored metallicity range, about 10$\times$ solar metallicity is required to reproduce the observed \ce{SO2} abundances. At solar metallicity or below, \ce{SO2} would not form at observable levels ($\sim$ 1 ppm), consistent with the metallicity constraint inferred from \ce{H2O}.   

\subsubsection{Tracing horizontal wind with \ce{CH4}}

\begin{figure}[t]
    \centering
    \includegraphics[width = 0.49\textwidth]{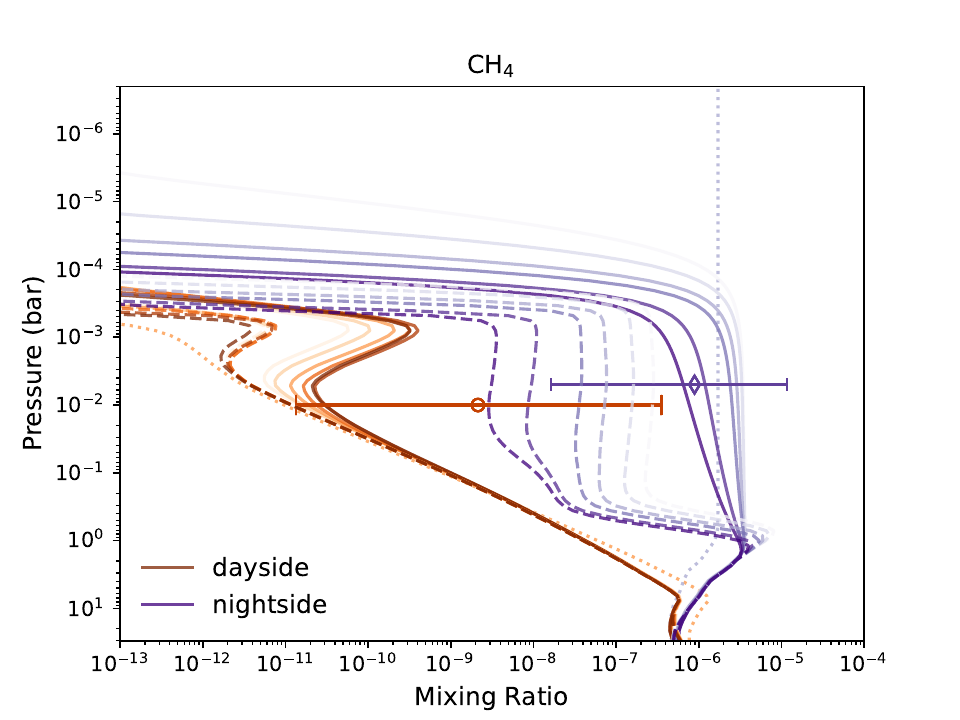}
    \caption{Same as Figure \ref{fig:so2_wind} but for \ce{CH4}. While the uncertainty on the retrieved \ce{CH4} VMRs is large, the nightside containing more \ce{CH4} is consistent with our predictions, particularly for all the strong vertical mixing cases or weak vertical mixing with zonal wind {$\lesssim$}3000 m/s.}
    \label{fig:ch4_wind}
\end{figure}

Among the retrieved molecules, \ce{CH4} exhibits the largest day-night variation (1-$\sigma$ $\log_{10}$VMR day: $-8.7^{+2.2}_{-2.2}$; night: $-6.1^{+1.1}_{-1.8}$). To further investigate the role of disequilibrium chemistry driven by global circulation, we again compare the retrieved \ce{CH4} abundances with those predicted by our two-column photochemical model. Unlike the photochemical product \ce{SO2}, \ce{CH4} is strongly regulated by transport-induced quenching \citep[e.g.,][]{Moses2011,Zamyatina2023,Lee2023}. \autoref{fig:ch4_wind} illustrates that the \ce{CH4} distribution is sensitive to both vertical and horizontal transport. Under weak vertical mixing (dashed lines in Figure \ref{fig:ch4_wind}), horizontal winds can efficiently transport \ce{CH4} from the nightside to the dayside, where it is subsequently destroyed. In contrast, under strong vertical mixing (solid lines in \autoref{fig:ch4_wind}), it dominates over horizontal transport, resulting in \ce{CH4} abundances that are largely insensitive to the wind speeds explored. 

Apart from the weak vertical mixing with strong zonal wind ($>$ 3000 m/s) scenario, most of the \ce{CH4} profiles are broadly consistent with the retrieved abundances, owing to the large observational uncertainties. Nevertheless, this analysis demonstrates the potential for simultaneously retrieving dayside and nightside \ce{SO2} and \ce{CH4} as tracers to constrain zonal wind in future observations.  Thus, future work may leverage \ce{SO2} to provide a lower bound on zonal wind speeds, while \ce{CH4} could provide an upper bound (assuming weak vertical mixing).

\subsubsection{The curious case of \ce{CO2}}

In principle, \ce{CO2} should play a similar role to \ce{H2O} as a tracer of atmospheric metallicity -- if not an even more sensitive tracer \citep{Zahnle2009b, Moses2013b,Wakeford2018}. However, the retrieved day and nightside \ce{CO2} abundances are difficult to reconcile in context with the other species. Furthermore, recent analysis of absorption asymmetries in the hot Jupiter WASP-39 b by \citet{Espinoza2024} also found a significantly weaker \ce{CO2} feature on the cooler morning limb.
While our retrieved nightside \ce{CO2} constraint is highly uncertain (even at $1\sigma$), the posterior peaks at a much lower abundance than found in the dayside retrievals adopted from \citet{wakeford2025_w17emission} as well as the WASP-17b NIRISS SOSS eclipse spectrum analysis reported by \citet{Gressier2025}. This lower abundance of \ce{CO2} on the nightside is difficult to explain, as \ce{CO2} is thermochemically favored at lower temperatures \citep[e.g.,][]{Moses2013b}. At the high temperatures on WASP-17 b, \ce{CO2} rapidly reaches thermochemical equilibrium with CO, and the contribution from OH oxidation of CO is negligible. \ce{CO2} generally does not exhibit notable vertical gradients, and is thought to be relatively unaffected by transport processes as well. If confirmed, these observations may suggest the presence of unknown mechanisms that reduces \ce{CO2} in the cooler conditions on the nightside.





\section{Discussion}
\label{sec:discuss}

We used the in-eclipse stellar spectrum of WASP-17 to isolate the day and nightside spectra of WASP-17b from the out-of-eclipse and out-of-transit baseline observations, respectively. Our results serve to validate the iESR PIE technique under specific stellar, planetary, and instrumental circumstances, while also hinting at novel transport-induced-chemistry on the nightside of WASP-17b.  

\subsection{On the Validity of iESR PIE}  
\label{sec:discuss:PIE} 

Our dayside PIE emission spectrum agrees well with the dayside eclipse spectrum extracted using a standard light curve fitting analysis, and the retrieval results generally agree with those performed on the full panchromatic WASP-17b dayside spectrum analyzed by \citet{wakeford2025_w17emission}. \autoref{fig:PIE1} shows that, while the dayside PIE spectra from the pre- and post-eclipse baseline are offset from one another, the offset is consistent with the known presence of a dayside hotspot offset eastward from the substellar point \citep{Valentine2024}, and the \eureka S6 emission spectrum lies between them, in good agreement with the structure in the spectrum. Retrievals on the dayside PIE spectrum allow two PT profile modes, one agrees with a panchromatic analysis of WASP-17b, while the other does not. Restricting the prior volume for our PT profile (using $\rm \log \gamma$) to explore and isolate the correct mode yields results that are in good agreement with the results obtained from the combined NIRISS SOSS, NIRSpec G395H, and MIRI LRS spectra \citep{wakeford2025_w17emission}. 
The nightside retrieval results favor an error inflation of $224\pm18$ ppm, which serves as a quantitative measure of the unaccounted for uncertainty due to changing astrophysical or instrumental characteristics between visits. As a result, the nightside atmospheric constraints are less precise than the dayside results and span a broad range of atmospheric parameters, thereby avoiding a bimodal solution for the PT profile as seen for the dayside by virtue of large uncertainties.   

Our novel nightside PIE results for WASP-17b can be explained by empirical trends and recent predictions from other hot Jupiters. 
The overall emission signal size from the nightside of WASP-17b corresponds to a temperature of around 1000 K. Such temperatures are consistent with constraints on the nightsides of hot Jupiters from phase curve measurements \citep[e.g.,][]{Beatty2019, Keating2019, Keating2020, Bell2024} and previous modeling results \citep[e.g.,][]{Kataria2016, Gao2021} that suggest uniform nightside temperatures near 1100 K for hot Jupiters nearly irrespective of the dayside temperature. The nightside temperature of WASP-17b agrees with these past findings, and this result serves as an initial line of evidence for the validity of the nightside emission extracted using the iERS PIE technique.  

The molecular absorption seen in the nightside spectrum and the retrieved gas abundances are consistent with day-night transport-induced chemistry modeling. The presence of abundant \ce{SO2} on the nightside of hot Jupiters was predicted by \citet{Tsai2023c} as part of the explanation for the observed terminator \ce{SO2} identified in the WASP-39b transmission spectrum \citep[][\citetalias{CarterMay2024} \citeyear{CarterMay2024}]{Alderson2023, Tsai2023a, Powell2024}. We demonstrated that the same mechanism can explain our tentative detection of \ce{SO2} on the nightside and the non-detection of \ce{SO2} on the dayside. Furthermore, the indication of \ce{CH4} on the nightside, but not the dayside, is also consistent with thermochemical expectations for \ce{CH4} to be the dominant carbon bearing species at the cooler temperatures found on the nightside. Taken together, these findings present a physically and chemically consistent explanation for the observed features in the nightside emission spectrum and offer an additional, non-trivial line of evidence toward the validation of the PIE technique as explored in this work. 

\subsection{On the Nightside of WASP-17b} 
\label{sec:discuss:nightside} 


Our novel data reduction methods used to extract and study the nightside spectrum of WASP-17b yielded compelling, if tentative, results that further our understanding of this benchmark hot Jupiter. We retrieve a nightside temperature of $\mathrm{T_{eq}} = 1005^{+129}_{-383}$ K. As discussed previously, this is consistent with other similar hot-Jupiters that have been found with nightside temperatures around 1000 K. Our necessary inclusion of error inflation in our nightside retrievals softens the statistical weight of our chemistry claims, leading to tentative detections of \ce{SO2} at $2.3\sigma$ ($\log \mathrm{SO_2}= -4.8^{+1.2}_{-1.2}$), \ce{CH4} at $1.5\sigma$ ($\log \mathrm{CH_4}=-6.1^{+1.1}_{-1.8}$), and \ce{H2O} at $1.5\sigma$ ($\log \mathrm{H_2O}=-4.1^{+1.8}_{-3.2}$). Despite the tentative detections, the broadly constrained abundances are consistent with nightside \ce{SO2} accumulation via horizontal transport from the dayside in a super-solar metallicity atmosphere at the temperatures probed on both the day and nightsides. 

\begin{figure}[t!]
    \centering
    \includegraphics[width=1.0\linewidth]{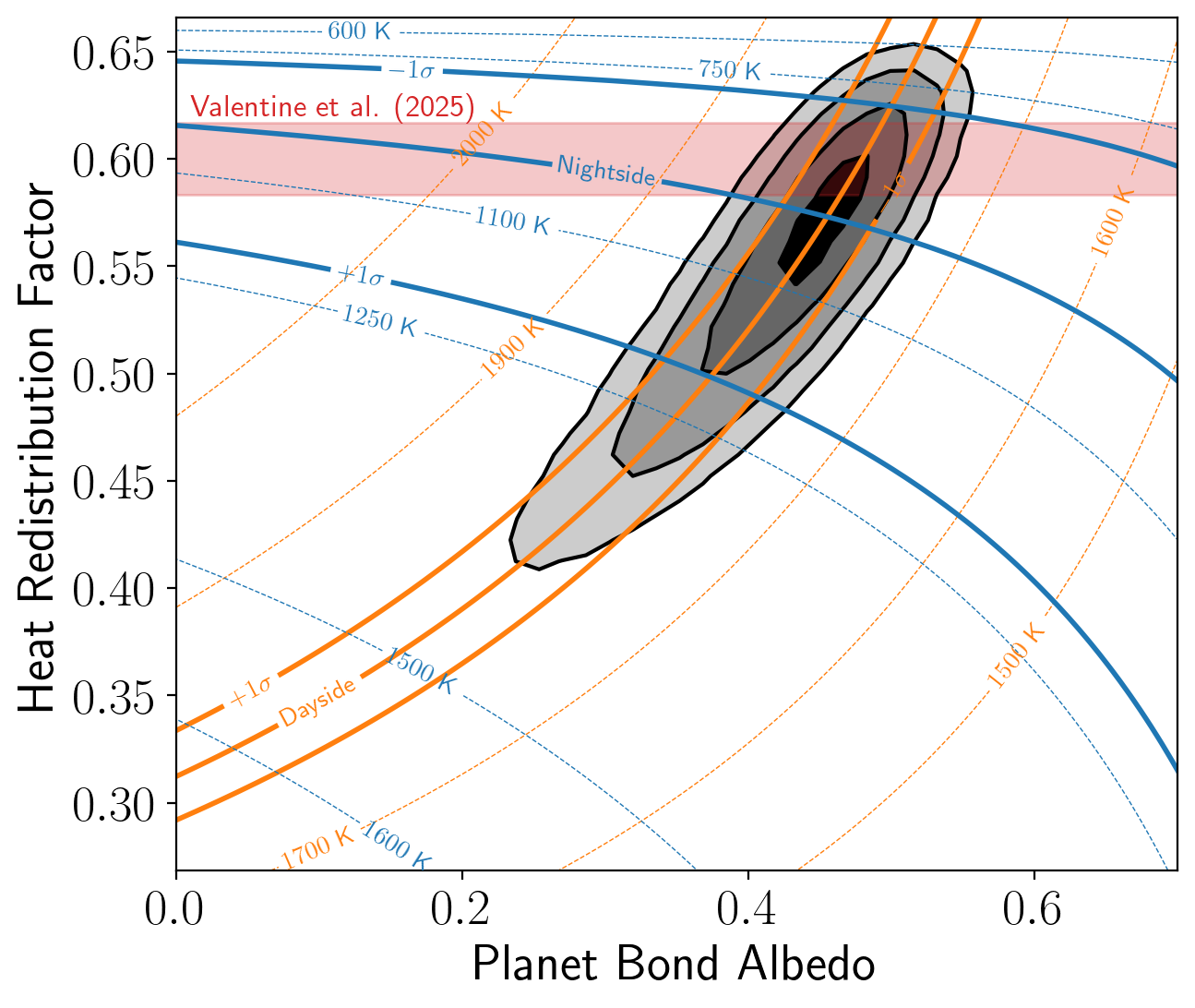}
    \caption{Our WASP-17b day and nightside temperature measurements constrain the heat redistribution ($f$) and bond albedo ($A_B$). Orange (blue) contours show possible dayside (nightside) equilibrium temperatures as a function of heat redistribution (y-axis) and bond albedo (x-axis). Thick contours show the median retrieved temperatures bounded by $\pm 1 \sigma$ uncertainties. Black contours show the joint constraints on $f$ and $A_B$ from our temperature measurements \citep[e.g., see Figure 1 in][{for an analogous example scenario}]{Cowan2011b}. Our nightside constraints agree with expectations based on the measured heat redistribution from the MIRI LRS eclipse map \citep{Valentine2024}.}
    \label{fig:daynight_contours}
\end{figure}

\autoref{fig:daynight_contours} places our dayside and, in particular, our novel nightside temperature measurements in context with first order energy balance parameters as well as previous independent WASP-17b findings. We use a Markov Chain Monte Carlo (MCMC; via \texttt{emcee} \citealp[]{emcee}) to sample heat redistribution factors ({$f \in [1/4,2/3]$;} {see \citealp{LopezMorales2007, Spiegel2010}}) and bond albedos ({$A_B \in [0,1]$}) subject to our NIRSpec G395H day and nightside equilibrium temperatures and associated uncertainties \citep[e.g., see][{their Figure 1 for an instructive example and discussion of joint constraints on circulation and bond albedo from day and night measurements}]{Cowan2011b}. We report 1D marginalized constraints $f = 0.54^{+0.06}_{-0.08}$ and $A_{B} = 0.42^{+0.06}_{-0.10}$ (at $1 \sigma$). Our measured nightside temperature and the corresponding constraint it places on heat redistribution agree very well with expectations based on the measured heat redistribution from the MIRI LRS eclipse map \citep{Valentine2024}\footnote{We note the linear transformation $g \rightarrow f$ from the redistribution efficiency of $g = 0.92\pm0.02$ reported in \cite{Valentine2024} to the redistribution factor $f = 0.6 \pm 0.02$ used here, as in \citet{Hansen2008} \citep[for more details, see][]{Spiegel2010, Morris2022}.}. Taken together, our nightside temperature results help to independently confirm that WASP-17b very likely has a nonzero bond albedo \citep{Gressier2025} and relatively inefficient heat redistribution to the nightside \citep{Valentine2024}. This indication of a nonzero bond albedo for WASP-17b is consistent with the tentative evidence of reflected light seen in the NIRISS SOSS dayside emission spectrum below 1 \micron{}, {which \citet{Gressier2025} report could be due to a geometric albedo of ${\sim}0.2$. Although our inferred Bond albedo of ${\sim}0.42$ is higher than the suggested geometric albedo, the factor of two discrepancy is consistent with the longstanding geometric-Bond albedo dichotomy seen in other hot Jupiters \citep[e.g.,][]{Schwartz2015, Crossfield2015b, Splinter2025}.}  The presence of \ce{SiO2(s)} clouds reported in the MIRI LRS transmission spectrum \citep{Grant2023} {provides additional circumstantial support for these high albedo claims.}   

Our inferred WASP-17b nightside characteristics are also in agreement with other recent findings on WASP-17b. Our constraints on the nightside \ce{H2O} and \ce{SO2} abundances are consistent with the supersolar metallicity for WASP-17b reported in \citet{Louie2024} and \citet{Gressier2025} from \ce{H2O} in the NIRISS transmission and emission spectra, respectively. The role of zonal winds in explaining the nightside \ce{SO2} is broadly consistent with the emission mapping results of \citet{Valentine2024} that showed a modest hotspot offset indicative of the presence of an equatorial jet, along with relatively inefficient heat redistribution from the dayside to the nightside. While we were not able to uniquely constrain the horizontal wind speed required to  produce the observed nightside \ce{SO2}, a large range spanning at least 100 m/s to 5 km/s is capable of doing so, consistent with the mean zonal wind speeds predicted by the GCM models of \citet{Kataria2016} ($\le 5$ km/s). Thus, even the inefficient heat transport suggested by \citet{Valentine2024} and our nightside temperature inferences are capable of producing detectable nightside \ce{SO2}. 








\subsection{Caveats, Challenges, \& Prospects for Future Work} 
\label{sec:discuss:caveats}

Much of this work has focused on the intriguing results that we obtained for the nightside of WASP-17b using NIRSpec G395H. However, the limitations of our methods that drove null results in our brief investigations using other instruments and data from other systems presents a caveat to this work and a challenge to future PIE observations.  

First, the iESR method that we used for the WASP-17b NIRSpec G395H data did not yield reliable nightside spectra for NIRISS SOSS or MIRI LRS observations of the same planet. Instead, as discussed in \autoref{sec:data:niriss_miri}, the systematics intrinsic to each instrument dominated the flux differences seen between eclipse and transit, rather than the change in planetary emission between the day and nightsides. Leveraging known instrument systematics from prior studies \citep[e.g.,][\citetalias{CarterMay2024} \citeyear{CarterMay2024}]{Holmberg2023, Dyrek2024} helped us identify these contaminating signals, but our preliminary attempts to remove them were unsuccessful and should be the subject of dedicated future work. At this time, we recommend that astronomers seeking to apply the iESR methods presented in this paper should use NIRSpec/G395H to observe transit and eclipse as close together {in time} as feasible. {This recommendation from our work with JWST may also apply to missions such as Ariel \citep{ARIEL2020} that could make use of multi-epoch phase curves, partial phase curves, and longer out-of-eclipse baselines \citep{CharnayAriel2022, Changeat2025}.} Moreover, instrument stability is likely to be an important requirement for any future mission seeking to leverage the PIE technique \citep[e.g.,][]{Mandell2022}.   

Second, even stable and reproducible transit and eclipse measurements obtained with NIRSpec G395H may not yield reliable nightside emission spectra using the iESR technique if the star is also changing on timescales comparable to the planet orbital period or the duration between separate observations. Our companion investigation into the HAT-P-26b {(a warm Neptune-like planet on a 4.23 day orbital period around a $\rm5079~K$ K-dwarf)} NIRSpec G395H observations presented in \citet{hatp26b_pie} demonstrates that iESR PIE for planets around later-type stars may be more challenging due to stellar variability between visits. {In that case, the HAT-P-26b transit and eclipse measurements were separated in time by 9.5 orbits or about 40.2 days. Although the rotation period of the star is not well constrained \citep{Hartman2011}, this long time gap between observations is sufficient to expect changes to have occurred.} Diagnosing such variability-driven effects required a stellar model-based correction, and correcting for them in the future will require improvements in these stellar models (and/or simultaneous access to broader and shorter wavelength data). These findings echo the challenges and needs of the JWST transiting exoplanet community, where transmission spectra of M-dwarf planets face stellar contamination from the TLS effect \citep[e.g.,][]{Lim2023, Moran2023, Canas2025}. These parallels are not simply illustrative; they are likely driven by the same effects: stellar spatial heterogeneities that exhibit unique spectroscopic signatures and which change between observing visits. Therefore, solutions to one or both of these problems may tread common ground. Future work should explore these techniques together as independent measures of the same stellar spatial, spectral, and temporal complexities. In the meantime, applications of iESR for other exoplanet targets should target adjacent transit and eclipse pairs, and separations of more than the star's rotation period should be avoided, if possible. {Ultimately, due to the instrumental and astrophysical caveats discussed above, future work should reanalyze a full/continuous phase curve to clarify remaining sources of uncertainty and further validate the iESR PIE technique.} 

Given that the study of stellar infrared excess originates in the disk community, how do we know the planetary signal is not dust? In this work, our method of iESR would divide out any static signals that are not time varying. For non-transiting PIE this would be of greater concern when directly fitting the calibrated combined light spectrum of flux sources from a given system. \citet{Lustig-Yaeger2021} used models to investigate confusion with warm disks for WASP-43b in theory and found that the presence of zodi should not bias inferred planet parameters from spectra containing both components. In practice, current and ongoing searches for planets around white dwarfs with JWST using PIE also contends with ambiguities due to dust \citep{Limbach2022, Limbach2024}.  \citet{Limbach2024} suggest that additional JWST observations could distinguish planet from dust by detecting or ruling out spectral features indicative of a planet atmosphere, similar to the nightside chemistry line of reasoning presented here. 


\section{Conclusions} 
\label{sec:conclusion}

We used in-eclipse JWST NIRSpec/G395H observations of WASP-17b to extract the WASP-17 stellar spectrum and then removed this contribution from the out-of-eclipse and out-of-transit baseline measurements. This enabled us to extract the dayside and nightside emission spectrum of WASP-17b from $2.9-5.2$ \microns{}. The dayside spectrum is in good agreement with traditional data reduction, light curve fitting techniques, and shows strong absorption from \ce{CO2}. The nightside spectrum is entirely novel and is consistent with temperatures around 1000 K and exhibits tentative evidence of absorption from \ce{SO2} and \ce{CH4}. In particular, the presence and loosely constrained \ce{SO2} abundance is consistent with transport-induced chemistry, thereby highlighting the unique capability of nightside measurements to probe critical transport processes. 

The method of in-eclipse spectrum removal that we used here for WASP-17b presents a compelling opportunity to reveal the nightsides of other hot exoplanets observed with JWST to achieve science previously only attainable with phase curves, but using a fraction of the observing time. 
However, our null result for HAT-P-26b \citep[see][]{hatp26b_pie} also demonstrated that using iESR for PIE to study smaller, cooler planets around time-variable stars may face significant practical challenges, requiring the development of extremely accurate stellar models.   
{These findings suggests that PIE is feasible with JWST/NIRSpec for two epochs separated in time by significantly less than the rotation period of the host star.}
The precision of JWST and the success of our applied methods to gain insight into the nightside of WASP-17b take a critical step forward for validating the PIE method {for multi-epoch phase measurements} on the path towards the spectroscopic characterization of non-transiting exoplanets in the mid-IR.  


\textbf{Acknowledgments}  {We express our gratitude to the anonymous reviewer whose comments improved the quality and clarity of this manuscript.} This paper reports work carried out in the context of the JWST Telescope Scientist Team with funding provided by the National Aeronautics and Space Administration (NASA) through Grant No. 80NSSC20K0586. This paper is based upon work supported by NASA through the Exoplanets Research Program (XRP) Grant No. 80NSSC23K0373. 
This material is also based upon work performed as part of the CHAMPs (Consortium on Habitability and Atmospheres of M-dwarf Planets) team, supported by NASA under Grant No. 80NSSC23K1399 issued through the Interdisciplinary Consortia for Astrobiology Research (ICAR) program. D.R.L. and S.P. acknowledge support from NASA under award number 80GSFC24M0006.


\software{\texttt{Astropy} \citep{astropy:2013, astropy:2018, astropy:2022}; \texttt{dynesty} \citep{Speagle2020}; \texttt{emcee} \citep{emcee}; {\eureka \citep{Bell2022};} \texttt{Matplotlib} \citep{Hunter2007}; \texttt{MultiNest} \citep{Feroz2009}; \texttt{PySynPhot} \citep{pysynphot, STScI2018}; \poseidon \citep{MacDonald2017, MacDonald2023}; \texttt{PyMultiNest} \citep{Buchner2014}}. 

%

\facilities{JWST(NIRSpec), Exo.Mast, NASA ADS}



\appendix

\section{Additional Retrieval Materials} 
\label{appendix:more_retrievals}

\begin{figure*}[t]
    \centering
    \includegraphics[width = 0.99\textwidth]{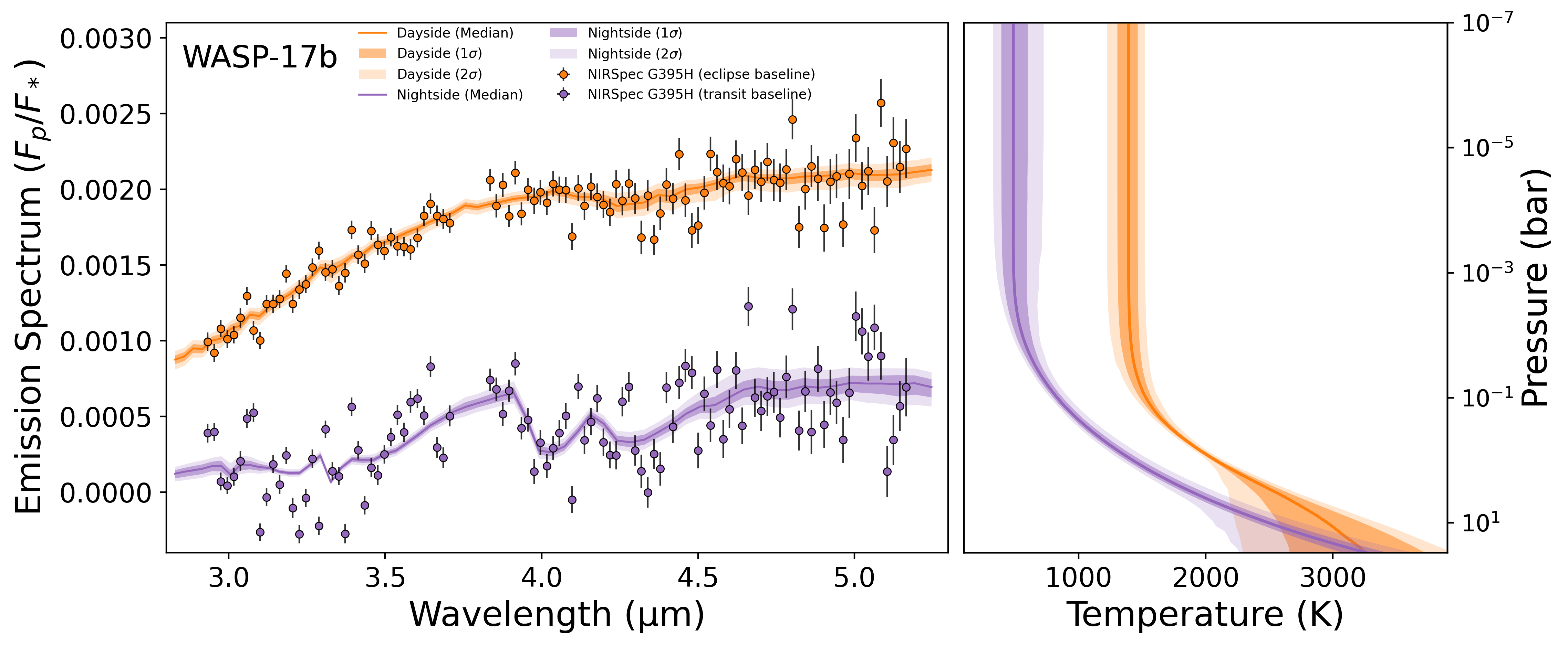}
    \includegraphics[width = 0.99\textwidth]{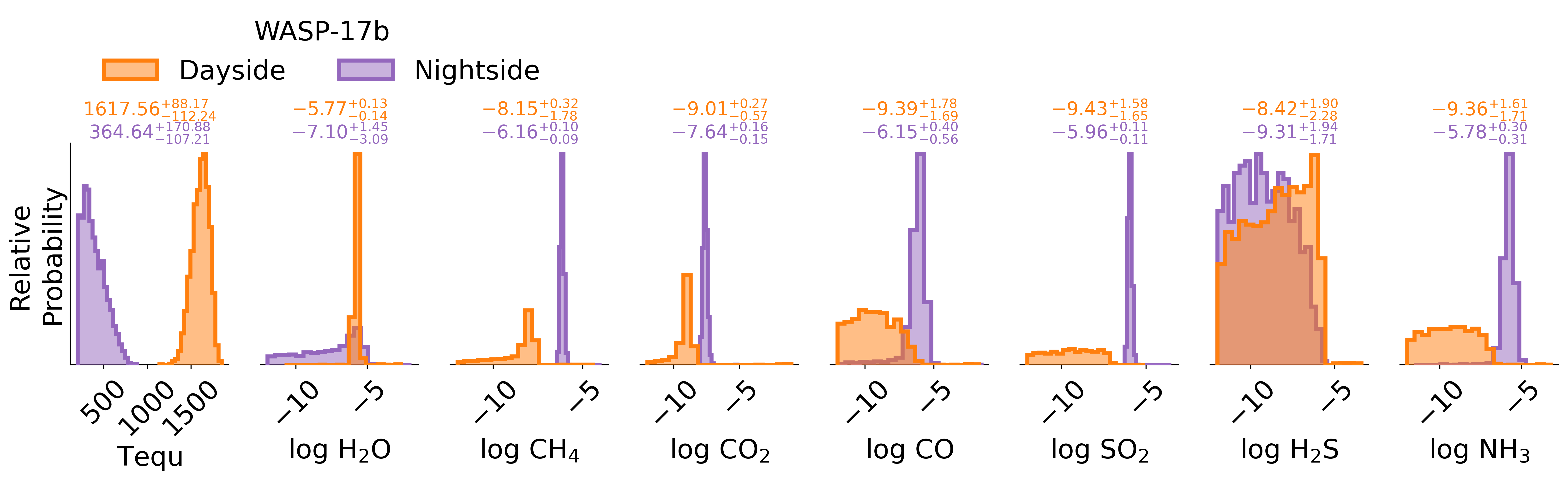}
    \caption{Similar to the retrieval results shown in \autoref{fig:retrieval2}, but without error inflation. The $\chi^2_{\nu}$ for the best-fitting spectral model is 1.9 and 8.3 for the dayside and nightside spectra, respectively.}
    \label{fig:retrieval2_no_infl}
\end{figure*}

\autoref{fig:retrieval2_no_infl} shows a summary of results from our WASP-17b PIE day and nightside retrievals, similar to \autoref{fig:retrieval2}, except now without the error inflation parameter. As shown in \autoref{tab:RetrievalResults}, these fits have higher reduced-$\chi^2$ statistics for the day and nightside spectra (1.8 and 8.3, respectively) and provide erroneously tight constraints on the retrieved planet parameters. Therefore, we favor the emission retrievals that use error inflation following the \citet{Line2015} prescription, particularly for reporting our nightside results.   

\autoref{tab:RetrievalResultsDayside} shows a series of retrieval results for the dayside emission spectrum of WASP-17b. Columns 2, 3 and 4 show NIRSpec G395H results with different priors on $\log \gamma$ to isolate the different solution modes. The last column shows results from \citet{wakeford2025_w17emission}, which exhibits good agreement with our ``high $\gamma$'' case despite our use of only NIRSpec data. We favor the more precise results from \citet{wakeford2025_w17emission} for our day-night chemical transport calculations in \autoref{sec:modeling:chemistry}. 

\begin{deluxetable*}{r||l|l|l|l}
\tablewidth{0.97\textwidth}
\tablecaption{Comparison of Dayside \poseidon Retrieval Model $1\sigma$ Posterior Constraints  \label{tab:RetrievalResultsDayside}} 
\tablehead{
\colhead{} & \multicolumn{3}{c}{NIRSpec G395H} & \multicolumn{1}{c}{NIRISS + NIRSpec + MIRI} \\
\cline{2-5} 
\colhead{Parameters} & \colhead{Broad $\gamma$} & \colhead{Low $\gamma$} & \colhead{High $\gamma$} & \colhead{Broad $\gamma$ \citep{wakeford2025_w17emission}}
}
\startdata
$\mathrm{R}_{\mathrm{p, \, ref}}$ & $1.699^{+0.018}_{-0.018}$ & $1.700^{+0.018}_{-0.017}$ & $1.696^{+0.017}_{-0.017}$ & $1.808^{+0.002}_{-0.002}$ \\
$\log \, \kappa\mathrm{IR}$ & $-4.79^{+0.20}_{-0.13}$ & $-4.81^{+0.17}_{-0.12}$ & $-3.08^{+0.25}_{-0.43}$ & $-2.55^{+0.60}_{-0.63}$ \\
$\log \, \gamma$ & $-1.26^{+0.35}_{-0.41}$ & $-1.36^{+0.37}_{-0.38}$ & $-0.250^{+0.030}_{-0.032}$ & $-0.391^{+0.035}_{-0.041}$ \\
$\mathrm{T_{int}}$ & $603^{+423}_{-357}$ & $603^{+353}_{-335}$ & $294^{+157}_{-161}$ & $196^{+179}_{-128}$ \\
$\mathrm{T_{equ}}$ & $1635^{+91}_{-106}$ & $1647^{+71}_{-98}$ & $1796^{+26}_{-28}$ & $1467^{+20}_{-12}$ \\
$\log \, \mathrm{H_2 O}$ & $-5.73^{+0.18}_{-0.20}$ & $-5.74^{+0.17}_{-0.17}$ & $-6.7^{+3.1}_{-3.6}$ & $-2.4^{+1.2}_{-0.7}$ \\
$\log \, \mathrm{CH_4}$ & $-9.1^{+1.1}_{-1.9}$ & $-9.1^{+1.1}_{-1.8}$ & $-5.3^{+0.6}_{-2.5}$ & $-8.7^{+2.2}_{-2.2}$ \\
$\log \, \mathrm{CO_2}$ & $-9.08^{+0.39}_{-1.12}$ & $-9.13^{+0.39}_{-1.22}$ & $-1.66^{+0.37}_{-0.71}$ & $-2.43^{+1.20}_{-0.79}$ \\
$\log \, \mathrm{CO}$ & $-9.1^{+1.8}_{-1.9}$ & $-9.3^{+1.7}_{-1.7}$ & $-6.4^{+3.4}_{-3.6}$ & $-6.5^{+3.7}_{-3.7}$ \\
$\log \, \mathrm{SO_2}$ & $-9.5^{+1.6}_{-1.6}$ & $-9.6^{+1.6}_{-1.5}$ & $-8.7^{+2.2}_{-2.2}$ & $-7.6^{+2.6}_{-2.9}$ \\
$\log \, \mathrm{H_2 S}$ & $-8.8^{+2.0}_{-2.1}$ & $-8.8^{+1.9}_{-2.0}$ & $-7.5^{+2.8}_{-2.8}$ & --- \\
$\log \, \mathrm{NH_3}$ & $-9.3^{+1.7}_{-1.7}$ & $-9.5^{+1.7}_{-1.6}$ & $-7.7^{+2.7}_{-2.7}$ & $-3.7^{+1.2}_{-0.8}$ \\
$\mathrm{b}$ & $-8.34^{+0.16}_{-0.17}$ & $-8.33^{+0.14}_{-0.14}$ & $-8.30^{+0.14}_{-0.16}$ & --- \\
\enddata
\end{deluxetable*}  

\bibliography{main}

\begin{thebibliography}{}
\expandafter\ifx\csname natexlab\endcsname\relax\def\natexlab#1{#1}\fi

\bibitem[{{Ahrer} {et~al.}(2023){Ahrer}, {Stevenson}, {Mansfield}, {Moran}, {Brande}, {Morello}, {Murray}, {Nikolov}, {Petit dit de la Roche}, {Schlawin}, {Wheatley}, {Zieba}, {Batalha}, {Damiano}, {Goyal}, {Lendl}, {Lothringer}, {Mukherjee}, {Ohno}, {Batalha}, {Battley}, {Bean}, {Beatty}, {Benneke}, {Berta-Thompson}, {Carter}, {Cubillos}, {Daylan}, {Espinoza}, {Gao}, {Gibson}, {Gill}, {Harrington}, {Hu}, {Kreidberg}, {Lewis}, {Line}, {L{\'o}pez-Morales}, {Parmentier}, {Powell}, {Sing}, {Tsai}, {Wakeford}, {Welbanks}, {Alam}, {Alderson}, {Allen}, {Anderson}, {Barstow}, {Bayliss}, {Bell}, {Blecic}, {Bryant}, {Burleigh}, {Carone}, {Casewell}, {Changeat}, {Chubb}, {Crossfield}, {Crouzet}, {Decin}, {D{\'e}sert}, {Feinstein}, {Flagg}, {Fortney}, {Gizis}, {Heng}, {Iro}, {Kempton}, {Kendrew}, {Kirk}, {Knutson}, {Komacek}, {Lagage}, {Leconte}, {Lustig-Yaeger}, {MacDonald}, {Mancini}, {May}, {Mayne}, {Miguel}, {Mikal-Evans}, {Molaverdikhani}, {Palle}, {Piaulet}, {Rackham}, {Redfield}, {Rogers}, {Roy}, {Rustamkulov},
  {Shkolnik}, {Sotzen}, {Taylor}, {Tremblin}, {Tucker}, {Turner}, {de Val-Borro}, {Venot}, \& {Zhang}}]{Ahrer2023}
{Ahrer}, E.-M., {Stevenson}, K.~B., {Mansfield}, M., {et~al.} 2023, Nature, 614, 653

\bibitem[{{Alderson} {et~al.}(2022){Alderson}, {Wakeford}, {MacDonald}, {Lewis}, {May}, {Grant}, {Sing}, {Stevenson}, {Fowler}, {Goyal}, {Batalha}, \& {Kataria}}]{Alderson2022}
{Alderson}, L., {Wakeford}, H.~R., {MacDonald}, R.~J., {et~al.} 2022, \mnras, 512, 4185

\bibitem[{{Alderson} {et~al.}(2023){Alderson}, {Wakeford}, {Alam}, {Batalha}, {Lothringer}, {Adams Redai}, {Barat}, {Brande}, {Damiano}, {Daylan}, {Espinoza}, {Flagg}, {Goyal}, {Grant}, {Hu}, {Inglis}, {Lee}, {Mikal-Evans}, {Ramos-Rosado}, {Roy}, {Wallack}, {Batalha}, {Bean}, {Benneke}, {Berta-Thompson}, {Carter}, {Changeat}, {Col{\'o}n}, {Crossfield}, {D{\'e}sert}, {Foreman-Mackey}, {Gibson}, {Kreidberg}, {Line}, {L{\'o}pez-Morales}, {Molaverdikhani}, {Moran}, {Morello}, {Moses}, {Mukherjee}, {Schlawin}, {Sing}, {Stevenson}, {Taylor}, {Aggarwal}, {Ahrer}, {Allen}, {Barstow}, {Bell}, {Blecic}, {Casewell}, {Chubb}, {Crouzet}, {Cubillos}, {Decin}, {Feinstein}, {Fortney}, {Harrington}, {Heng}, {Iro}, {Kempton}, {Kirk}, {Knutson}, {Krick}, {Leconte}, {Lendl}, {MacDonald}, {Mancini}, {Mansfield}, {May}, {Mayne}, {Miguel}, {Nikolov}, {Ohno}, {Palle}, {Parmentier}, {Petit dit de la Roche}, {Piaulet}, {Powell}, {Rackham}, {Redfield}, {Rogers}, {Rustamkulov}, {Tan}, {Tremblin}, {Tsai}, {Turner}, {de Val-Borro},
  {Venot}, {Welbanks}, {Wheatley}, \& {Zhang}}]{Alderson2023}
{Alderson}, L., {Wakeford}, H.~R., {Alam}, M.~K., {et~al.} 2023, \nat, 614, 664

\bibitem[{{Allard} {et~al.}(2012){Allard}, {Homeier}, \& {Freytag}}]{Allard2012}
{Allard}, F., {Homeier}, D., \& {Freytag}, B. 2012, Philosophical Transactions of the Royal Society of London Series A, 370, 2765

\bibitem[{{Anderson} {et~al.}(2010){Anderson}, {Hellier}, {Gillon}, {Triaud}, {Smalley}, {Hebb}, {Collier Cameron}, {Maxted}, {Queloz}, {West}, {Bentley}, {Enoch}, {Horne}, {Lister}, {Mayor}, {Parley}, {Pepe}, {Pollacco}, {S{\'e}gransan}, {Udry}, \& {Wilson}}]{Anderson2010}
{Anderson}, D.~R., {Hellier}, C., {Gillon}, M., {et~al.} 2010, \apj, 709, 159

\bibitem[{{Anderson} {et~al.}(2011){Anderson}, {Smith}, {Lanotte}, {Barman}, {Collier Cameron}, {Campo}, {Gillon}, {Harrington}, {Hellier}, {Maxted}, {Queloz}, {Triaud}, \& {Wheatley}}]{Anderson2011}
{Anderson}, D.~R., {Smith}, A.~M.~S., {Lanotte}, A.~A., {et~al.} 2011, \mnras, 416, 2108

\bibitem[{{Arcangeli} {et~al.}(2021){Arcangeli}, {D{\'e}sert}, {Parmentier}, {Tsai}, \& {Stevenson}}]{Arcangeli2021}
{Arcangeli}, J., {D{\'e}sert}, J.~M., {Parmentier}, V., {Tsai}, S.~M., \& {Stevenson}, K.~B. 2021, \aap, 646, A94

\bibitem[{{Astropy Collaboration} {et~al.}(2013){Astropy Collaboration}, {Robitaille}, {Tollerud}, {Greenfield}, {Droettboom}, {Bray}, {Aldcroft}, {Davis}, {Ginsburg}, {Price-Whelan}, {Kerzendorf}, {Conley}, {Crighton}, {Barbary}, {Muna}, {Ferguson}, {Grollier}, {Parikh}, {Nair}, {Unther}, {Deil}, {Woillez}, {Conseil}, {Kramer}, {Turner}, {Singer}, {Fox}, {Weaver}, {Zabalza}, {Edwards}, {Azalee Bostroem}, {Burke}, {Casey}, {Crawford}, {Dencheva}, {Ely}, {Jenness}, {Labrie}, {Lim}, {Pierfederici}, {Pontzen}, {Ptak}, {Refsdal}, {Servillat}, \& {Streicher}}]{astropy:2013}
{Astropy Collaboration}, {Robitaille}, T.~P., {Tollerud}, E.~J., {et~al.} 2013, \aap, 558, A33

\bibitem[{{Astropy Collaboration} {et~al.}(2018){Astropy Collaboration}, {Price-Whelan}, {Sip{\H{o}}cz}, {G{\"u}nther}, {Lim}, {Crawford}, {Conseil}, {Shupe}, {Craig}, {Dencheva}, {Ginsburg}, {Vand erPlas}, {Bradley}, {P{\'e}rez-Su{\'a}rez}, {de Val-Borro}, {Aldcroft}, {Cruz}, {Robitaille}, {Tollerud}, {Ardelean}, {Babej}, {Bach}, {Bachetti}, {Bakanov}, {Bamford}, {Barentsen}, {Barmby}, {Baumbach}, {Berry}, {Biscani}, {Boquien}, {Bostroem}, {Bouma}, {Brammer}, {Bray}, {Breytenbach}, {Buddelmeijer}, {Burke}, {Calderone}, {Cano Rodr{\'\i}guez}, {Cara}, {Cardoso}, {Cheedella}, {Copin}, {Corrales}, {Crichton}, {D'Avella}, {Deil}, {Depagne}, {Dietrich}, {Donath}, {Droettboom}, {Earl}, {Erben}, {Fabbro}, {Ferreira}, {Finethy}, {Fox}, {Garrison}, {Gibbons}, {Goldstein}, {Gommers}, {Greco}, {Greenfield}, {Groener}, {Grollier}, {Hagen}, {Hirst}, {Homeier}, {Horton}, {Hosseinzadeh}, {Hu}, {Hunkeler}, {Ivezi{\'c}}, {Jain}, {Jenness}, {Kanarek}, {Kendrew}, {Kern}, {Kerzendorf}, {Khvalko}, {King}, {Kirkby}, {Kulkarni},
  {Kumar}, {Lee}, {Lenz}, {Littlefair}, {Ma}, {Macleod}, {Mastropietro}, {McCully}, {Montagnac}, {Morris}, {Mueller}, {Mumford}, {Muna}, {Murphy}, {Nelson}, {Nguyen}, {Ninan}, {N{\"o}the}, {Ogaz}, {Oh}, {Parejko}, {Parley}, {Pascual}, {Patil}, {Patil}, {Plunkett}, {Prochaska}, {Rastogi}, {Reddy Janga}, {Sabater}, {Sakurikar}, {Seifert}, {Sherbert}, {Sherwood-Taylor}, {Shih}, {Sick}, {Silbiger}, {Singanamalla}, {Singer}, {Sladen}, {Sooley}, {Sornarajah}, {Streicher}, {Teuben}, {Thomas}, {Tremblay}, {Turner}, {Terr{\'o}n}, {van Kerkwijk}, {de la Vega}, {Watkins}, {Weaver}, {Whitmore}, {Woillez}, {Zabalza}, \& {Astropy Contributors}}]{astropy:2018}
{Astropy Collaboration}, {Price-Whelan}, A.~M., {Sip{\H{o}}cz}, B.~M., {et~al.} 2018, \aj, 156, 123

\bibitem[{{Astropy Collaboration} {et~al.}(2022){Astropy Collaboration}, {Price-Whelan}, {Lim}, {Earl}, {Starkman}, {Bradley}, {Shupe}, {Patil}, {Corrales}, {Brasseur}, {N{"o}the}, {Donath}, {Tollerud}, {Morris}, {Ginsburg}, {Vaher}, {Weaver}, {Tocknell}, {Jamieson}, {van Kerkwijk}, {Robitaille}, {Merry}, {Bachetti}, {G{"u}nther}, {Aldcroft}, {Alvarado-Montes}, {Archibald}, {B{'o}di}, {Bapat}, {Barentsen}, {Baz{'a}n}, {Biswas}, {Boquien}, {Burke}, {Cara}, {Cara}, {Conroy}, {Conseil}, {Craig}, {Cross}, {Cruz}, {D'Eugenio}, {Dencheva}, {Devillepoix}, {Dietrich}, {Eigenbrot}, {Erben}, {Ferreira}, {Foreman-Mackey}, {Fox}, {Freij}, {Garg}, {Geda}, {Glattly}, {Gondhalekar}, {Gordon}, {Grant}, {Greenfield}, {Groener}, {Guest}, {Gurovich}, {Handberg}, {Hart}, {Hatfield-Dodds}, {Homeier}, {Hosseinzadeh}, {Jenness}, {Jones}, {Joseph}, {Kalmbach}, {Karamehmetoglu}, {Ka{l}uszy{'n}ski}, {Kelley}, {Kern}, {Kerzendorf}, {Koch}, {Kulumani}, {Lee}, {Ly}, {Ma}, {MacBride}, {Maljaars}, {Muna}, {Murphy}, {Norman}, {O'Steen},
  {Oman}, {Pacifici}, {Pascual}, {Pascual-Granado}, {Patil}, {Perren}, {Pickering}, {Rastogi}, {Roulston}, {Ryan}, {Rykoff}, {Sabater}, {Sakurikar}, {Salgado}, {Sanghi}, {Saunders}, {Savchenko}, {Schwardt}, {Seifert-Eckert}, {Shih}, {Jain}, {Shukla}, {Sick}, {Simpson}, {Singanamalla}, {Singer}, {Singhal}, {Sinha}, {Sip{H{o}}cz}, {Spitler}, {Stansby}, {Streicher}, {{{S}}umak}, {Swinbank}, {Taranu}, {Tewary}, {Tremblay}, {Val-Borro}, {Van Kooten}, {Vasovi{'c}}, {Verma}, {de Miranda Cardoso}, {Williams}, {Wilson}, {Winkel}, {Wood-Vasey}, {Xue}, {Yoachim}, {Zhang}, {Zonca}, \& {Astropy Project Contributors}}]{astropy:2022}
{Astropy Collaboration}, {Price-Whelan}, A.~M., {Lim}, P.~L., {et~al.} 2022, \apj, 935, 167

\bibitem[{{Azzam} {et~al.}(2016){Azzam}, {Tennyson}, {Yurchenko}, \& {Naumenko}}]{Azzam2016}
{Azzam}, A. A.~A., {Tennyson}, J., {Yurchenko}, S.~N., \& {Naumenko}, O.~V. 2016, \mnras, 460, 4063

\bibitem[{{Beatty} {et~al.}(2019){Beatty}, {Marley}, {Gaudi}, {Col{\'o}n}, {Fortney}, \& {Showman}}]{Beatty2019}
{Beatty}, T.~G., {Marley}, M.~S., {Gaudi}, B.~S., {et~al.} 2019, \aj, 158, 166

\bibitem[{{Bell} {et~al.}(2022){Bell}, {Ahrer}, {Brande}, {Carter}, {Feinstein}, {Caloca}, {Mansfield}, {Zieba}, {Piaulet}, {Benneke}, {Filippazzo}, {May}, {Roy}, {Kreidberg}, \& {Stevenson}}]{Bell2022}
{Bell}, T., {Ahrer}, E.-M., {Brande}, J., {et~al.} 2022, The Journal of Open Source Software, 7, 4503

\bibitem[{{Bell} {et~al.}(2024){Bell}, {Crouzet}, {Cubillos}, {Kreidberg}, {Piette}, {Roman}, {Barstow}, {Blecic}, {Carone}, {Coulombe}, {Ducrot}, {Hammond}, {Mendon{\c{c}}a}, {Moses}, {Parmentier}, {Stevenson}, {Teinturier}, {Zhang}, {Batalha}, {Bean}, {Benneke}, {Charnay}, {Chubb}, {Demory}, {Gao}, {Lee}, {L{\'o}pez-Morales}, {Morello}, {Rauscher}, {Sing}, {Tan}, {Venot}, {Wakeford}, {Aggarwal}, {Ahrer}, {Alam}, {Baeyens}, {Barrado}, {Caceres}, {Carter}, {Casewell}, {Challener}, {Crossfield}, {Decin}, {D{\'e}sert}, {Dobbs-Dixon}, {Dyrek}, {Espinoza}, {Feinstein}, {Gibson}, {Harrington}, {Helling}, {Hu}, {Iro}, {Kempton}, {Kendrew}, {Komacek}, {Krick}, {Lagage}, {Leconte}, {Lendl}, {Lewis}, {Lothringer}, {Malsky}, {Mancini}, {Mansfield}, {Mayne}, {Evans-Soma}, {Molaverdikhani}, {Nikolov}, {Nixon}, {Palle}, {Petit dit de la Roche}, {Piaulet}, {Powell}, {Rackham}, {Schneider}, {Steinrueck}, {Taylor}, {Welbanks}, {Yurchenko}, {Zhang}, \& {Zieba}}]{Bell2024}
{Bell}, T.~J., {Crouzet}, N., {Cubillos}, P.~E., {et~al.} 2024, Nature Astronomy, 8, 879

\bibitem[{{Buchner} {et~al.}(2014){Buchner}, {Georgakakis}, {Nandra}, {Hsu}, {Rangel}, {Brightman}, {Merloni}, {Salvato}, {Donley}, \& {Kocevski}}]{Buchner2014}
{Buchner}, J., {Georgakakis}, A., {Nandra}, K., {et~al.} 2014, \aap, 564, A125

\bibitem[{{Ca{\~n}as} {et~al.}(2025){Ca{\~n}as}, {Lustig-Yaeger}, {Tsai}, {M{\"u}ller}, {Helled}, {Louie}, {Guzm{\'a}n Caloca}, {Kanodia}, {Gao}, {Libby-Roberts}, {Hardegree-Ullman}, {Col{\'o}n}, {Czekala}, {Delamer}, {Han}, {Lin}, {Mahadevan}, {May}, {Ninan}, {Piette}, {Stef{\'a}nsson}, {Stevenson}, {Teske}, \& {Wallack}}]{Canas2025}
{Ca{\~n}as}, C.~I., {Lustig-Yaeger}, J., {Tsai}, S.-M., {et~al.} 2025, arXiv e-prints, arXiv:2502.06966

\bibitem[{{Cadieux} {et~al.}(2024){Cadieux}, {Doyon}, {MacDonald}, {Turbet}, {Artigau}, {Lim}, {Radica}, {Fauchez}, {Salhi}, {Dang}, {Albert}, {Coulombe}, {Cowan}, {Lafreni{\`e}re}, {L'Heureux}, {Piaulet-Ghorayeb}, {Benneke}, {Cloutier}, {Charnay}, {Cook}, {Fournier-Tondreau}, {Plotnykov}, \& {Valencia}}]{Cadieux2024}
{Cadieux}, C., {Doyon}, R., {MacDonald}, R.~J., {et~al.} 2024, \apjl, 970, L2

\bibitem[{{Carter} {et~al.}(2024){Carter}, {May}, {Espinoza}, {Welbanks}, {Ahrer}, {Alderson}, {Brahm}, {Feinstein}, {Grant}, {Line}, {Morello}, {O'Steen}, {Radica}, {Rustamkulov}, {Stevenson}, {Turner}, {Alam}, {Anderson}, {Batalha}, {Battley}, {Bayliss}, {Bean}, {Benneke}, {Berta-Thompson}, {Brande}, {Bryant}, {Burleigh}, {Coulombe}, {Crossfield}, {Damiano}, {D{\'e}sert}, {Flagg}, {Gill}, {Inglis}, {Kirk}, {Knutson}, {Kreidberg}, {L{\'o}pez Morales}, {Mansfield}, {Moran}, {Murray}, {Nixon}, {Petit dit de la Roche}, {Rackham}, {Schlawin}, {Sing}, {Wakeford}, {Wallack}, {Wheatley}, {Zieba}, {Aggarwal}, {Barstow}, {Bell}, {Blecic}, {Caceres}, {Crouzet}, {Cubillos}, {Daylan}, {de Val-Borro}, {Decin}, {Fortney}, {Gibson}, {Heng}, {Hu}, {Kempton}, {Lagage}, {Lothringer}, {Lustig-Yaeger}, {Mancini}, {Mayne}, {Mayorga}, {Molaverdikhani}, {Nasedkin}, {Ohno}, {Parmentier}, {Powell}, {Redfield}, {Roy}, {Taylor}, \& {Zhang}}]{CarterMay2024}
{Carter}, A.~L., {May}, E.~M., {Espinoza}, N., {et~al.} 2024, Nature Astronomy, 8, 1008

\bibitem[{{Changeat} {et~al.}(2025){Changeat}, {Lagage}, {Tinetti}, {Charnay}, {Cowan}, {Danielski}, {Ducrot}, {Dyrek}, {Edwards}, {Lueftinger}, {Micela}, {Morello}, {Pascale}, {Robert}, {Venot}, {Barstow}, {Bocchieri}, {Y-K. Cho}, {Cloutier}, {Coustenis}, {Lavvas}, {Miguel}, \& {Hou Yip}}]{Changeat2025}
{Changeat}, Q., {Lagage}, P.-O., {Tinetti}, G., {et~al.} 2025, arXiv e-prints, arXiv:2509.02657

\bibitem[{{Charnay} {et~al.}(2022){Charnay}, {Mendon{\c{c}}a}, {Kreidberg}, {Cowan}, {Taylor}, {Bell}, {Demangeon}, {Edwards}, {Haswell}, {Morello}, {Mugnai}, {Pascale}, {Tinetti}, {Tremblin}, \& {Zellem}}]{CharnayAriel2022}
{Charnay}, B., {Mendon{\c{c}}a}, J.~M., {Kreidberg}, L., {et~al.} 2022, Experimental Astronomy, 53, 417

\bibitem[{{Coles} {et~al.}(2019){Coles}, {Yurchenko}, \& {Tennyson}}]{Coles2019}
{Coles}, P.~A., {Yurchenko}, S.~N., \& {Tennyson}, J. 2019, \mnras, 490, 4638

\bibitem[{{Coulombe} {et~al.}(2023){Coulombe}, {Benneke}, {Challener}, {Piette}, {Wiser}, {Mansfield}, {MacDonald}, {Beltz}, {Feinstein}, {Radica}, {Savel}, {Dos Santos}, {Bean}, {Parmentier}, {Wong}, {Rauscher}, {Komacek}, {Kempton}, {Tan}, {Hammond}, {Lewis}, {Line}, {Lee}, {Shivkumar}, {Crossfield}, {Nixon}, {Rackham}, {Wakeford}, {Welbanks}, {Zhang}, {Batalha}, {Berta-Thompson}, {Changeat}, {D{\'e}sert}, {Espinoza}, {Goyal}, {Harrington}, {Knutson}, {Kreidberg}, {L{\'o}pez-Morales}, {Shporer}, {Sing}, {Stevenson}, {Aggarwal}, {Ahrer}, {Alam}, {Bell}, {Blecic}, {Caceres}, {Carter}, {Casewell}, {Crouzet}, {Cubillos}, {Decin}, {Fortney}, {Gibson}, {Heng}, {Henning}, {Iro}, {Kendrew}, {Lagage}, {Leconte}, {Lendl}, {Lothringer}, {Mancini}, {Mikal-Evans}, {Molaverdikhani}, {Nikolov}, {Ohno}, {Palle}, {Piaulet}, {Redfield}, {Roy}, {Tsai}, {Venot}, \& {Wheatley}}]{Coulombe2023}
{Coulombe}, L.-P., {Benneke}, B., {Challener}, R., {et~al.} 2023, \nat, 620, 292

\bibitem[{{Cowan} \& {Agol}(2011)}]{Cowan2011b}
{Cowan}, N.~B., \& {Agol}, E. 2011, \apj, 729, 54

\bibitem[{{Cowan} {et~al.}(2007){Cowan}, {Agol}, \& {Charbonneau}}]{Cowan2007}
{Cowan}, N.~B., {Agol}, E., \& {Charbonneau}, D. 2007, \mnras, 379, 641

\bibitem[{{Crossfield}(2015)}]{Crossfield2015b}
{Crossfield}, I.~J.~M. 2015, \pasp, 127, 941

\bibitem[{{Crossfield} {et~al.}(2010){Crossfield}, {Hansen}, {Harrington}, {Cho}, {Deming}, {Menou}, \& {Seager}}]{Crossfield2010}
{Crossfield}, I. J.~M., {Hansen}, B. M.~S., {Harrington}, J., {et~al.} 2010, \apj, 723, 1436

\bibitem[{{Doyon} {et~al.}(2023){Doyon}, {Willott}, {Hutchings}, {Sivaramakrishnan}, {Albert}, {Lafreni{\`e}re}, {Rowlands}, {Bego{\~n}a Vila}, {Martel}, {LaMassa}, {Aldridge}, {Artigau}, {Cameron}, {Chayer}, {Cook}, {Cooper}, {Darveau-Bernier}, {Dupuis}, {Earnshaw}, {Espinoza}, {Filippazzo}, {Fullerton}, {Gaudreau}, {Gawlik}, {Goudfrooij}, {Haley}, {Kammerer}, {Kendall}, {Lambros}, {Ignat}, {Maszkiewicz}, {McColgan}, {Morishita}, {Ouellette}, {Pacifici}, {Philippi}, {Radica}, {Ravindranath}, {Rowe}, {Roy}, {Roy}, {Saad}, {Sohn}, {Talens}, {Touahri}, {Thatte}, {Taylor}, {Vandal}, {Volk}, {Wander}, {Warner}, {Zheng}, {Zhou}, {Abraham}, {Beaulieu}, {Benneke}, {Ferrarese}, {Jayawardhana}, {Johnstone}, {Kaltenegger}, {Meyer}, {Pipher}, {Rameau}, {Rieke}, {Salhi}, \& {Sawicki}}]{Doyon2023}
{Doyon}, R., {Willott}, C.~J., {Hutchings}, J.~B., {et~al.} 2023, \pasp, 135, 098001

\bibitem[{{Dyrek} {et~al.}(2024){Dyrek}, {Ducrot}, {Lagage}, {Tremblin}, {Kendrew}, {Bouwman}, \& {Bouffet}}]{Dyrek2024}
{Dyrek}, A., {Ducrot}, E., {Lagage}, P.~O., {et~al.} 2024, \aap, 683, A212

\bibitem[{{Espinoza} {et~al.}(2024){Espinoza}, {Steinrueck}, {Kirk}, {MacDonald}, {Savel}, {Arnold}, {Kempton}, {Murphy}, {Carone}, {Zamyatina}, {Lewis}, {Samra}, {Kiefer}, {Rauscher}, {Christie}, {Mayne}, {Helling}, {Rustamkulov}, {Parmentier}, {May}, {Carter}, {Zhang}, {L{\'o}pez-Morales}, {Allen}, {Blecic}, {Decin}, {Mancini}, {Molaverdikhani}, {Rackham}, {Palle}, {Tsai}, {Ahrer}, {Bean}, {Crossfield}, {Haegele}, {H{\'e}brard}, {Kreidberg}, {Powell}, {Schneider}, {Welbanks}, {Wheatley}, {Brahm}, \& {Crouzet}}]{Espinoza2024}
{Espinoza}, N., {Steinrueck}, M.~E., {Kirk}, J., {et~al.} 2024, \nat, 632, 1017

\bibitem[{{Fauchez} {et~al.}(2025){Fauchez}, {Ducrot}, {Rackham}, {Stevenson}, {Mayorga}, \& {de Wit}}]{Fauchez2025}
{Fauchez}, T.~J., {Ducrot}, E., {Rackham}, B.~V., {et~al.} 2025, arXiv e-prints, arXiv:2502.19585

\bibitem[{{Feroz} {et~al.}(2009){Feroz}, {Hobson}, \& {Bridges}}]{Feroz2009}
{Feroz}, F., {Hobson}, M.~P., \& {Bridges}, M. 2009, \mnras, 398, 1601

\bibitem[{{Foreman-Mackey} {et~al.}(2013){Foreman-Mackey}, {Hogg}, {Lang}, \& {Goodman}}]{emcee}
{Foreman-Mackey}, D., {Hogg}, D.~W., {Lang}, D., \& {Goodman}, J. 2013, \pasp, 125, 306

\bibitem[{{Fortune} {et~al.}(2025){Fortune}, {Gibson}, {Diamond-Lowe}, {Mendon{\c{c}}a}, {Gressier}, {Kitzmann}, {Allen}, {August}, {Ih}, {Meier Vald{\'e}s}, {Zgraggen}, {Buchhave}, {Demory}, {Espinoza}, {Heng}, {Jones}, \& {Rathcke}}]{Fortune2025}
{Fortune}, M., {Gibson}, N.~P., {Diamond-Lowe}, H., {et~al.} 2025, arXiv e-prints, arXiv:2505.22186

\bibitem[{{Gandhi} \& {Madhusudhan}(2018)}]{Gandhi2018}
{Gandhi}, S., \& {Madhusudhan}, N. 2018, \mnras, 474, 271

\bibitem[{{Gao} \& {Powell}(2021)}]{Gao2021}
{Gao}, P., \& {Powell}, D. 2021, \apjl, 918, L7

\bibitem[{{Garcia} {et~al.}(2022){Garcia}, {Moran}, {Rackham}, {Wakeford}, {Gillon}, {de Wit}, \& {Lewis}}]{Garcia2022}
{Garcia}, L.~J., {Moran}, S.~E., {Rackham}, B.~V., {et~al.} 2022, \aap, 665, A19

\bibitem[{{Grant} {et~al.}(2023){Grant}, {Lewis}, {Wakeford}, {Batalha}, {Glidden}, {Goyal}, {Mullens}, {MacDonald}, {May}, {Seager}, {Stevenson}, {Valenti}, {Visscher}, {Alderson}, {Allen}, {Ca{\~n}as}, {Col{\'o}n}, {Clampin}, {Espinoza}, {Gressier}, {Huang}, {Lin}, {Long}, {Louie}, {Pe{\~n}a-Guerrero}, {Ranjan}, {Sotzen}, {Valentine}, {Anderson}, {Balmer}, {Bellini}, {Hoch}, {Kammerer}, {Libralato}, {Mountain}, {Perrin}, {Pueyo}, {Rickman}, {Rebollido}, {Sohn}, {van der Marel}, \& {Watkins}}]{Grant2023}
{Grant}, D., {Lewis}, N.~K., {Wakeford}, H.~R., {et~al.} 2023, \apjl, 956, L32

\bibitem[{{Gressier} {et~al.}(2025){Gressier}, {MacDonald}, {Espinoza}, {Wakeford}, {Lewis}, {Goyal}, {Louie}, {Radica}, {Batalha}, {Long}, {May}, {Mullens}, {Seager}, {Stevenson}, {Valenti}, {Alderson}, {Allen}, {Ca{\~n}as}, {Challener}, {Col{\'o}n}, {Glidden}, {Grant}, {Huang}, {Lin}, {Valentine}, {Mountain}, {Pueyo}, {Perrin}, \& {van der Marel}}]{Gressier2025}
{Gressier}, A., {MacDonald}, R.~J., {Espinoza}, N., {et~al.} 2025, \aj, 169, 57

\bibitem[{{Guillot}(2010)}]{Guillot2010}
{Guillot}, T. 2010, \aap, 520, A27

\bibitem[{{Hansen}(2008)}]{Hansen2008}
{Hansen}, B. M.~S. 2008, \apjs, 179, 484

\bibitem[{{\jhauth{Harrington}} {et~al.}(2006)}]{HarringtonEtal2006sciuandbphas}
{\jhauth{Harrington}}, {\jhauth{J}}., {et~al.} 2006, Science, 314, 623

\bibitem[{{Hartman} {et~al.}(2011){Hartman}, {Bakos}, {Kipping}, {Torres}, {Kov{\'a}cs}, {Noyes}, {Latham}, {Howard}, {Fischer}, {Johnson}, {Marcy}, {Isaacson}, {Quinn}, {Buchhave}, {B{\'e}ky}, {Sasselov}, {Stefanik}, {Esquerdo}, {Everett}, {Perumpilly}, {L{\'a}z{\'a}r}, {Papp}, \& {S{\'a}ri}}]{Hartman2011}
{Hartman}, J.~D., {Bakos}, G.~{\'A}., {Kipping}, D.~M., {et~al.} 2011, \apj, 728, 138

\bibitem[{{Holmberg} \& {Madhusudhan}(2023)}]{Holmberg2023}
{Holmberg}, M., \& {Madhusudhan}, N. 2023, \mnras, 524, 377

\bibitem[{{Horne}(1986)}]{Horne1986}
{Horne}, K. 1986, Publ. Astron. Soc. Pac., 98, 609

\bibitem[{Hunter(2007)}]{Hunter2007}
Hunter, J.~D. 2007, Computing in Science \& Engineering, 9, 90

\bibitem[{{Husser} {et~al.}(2013){Husser}, {Wende-von Berg}, {Dreizler}, {Homeier}, {Reiners}, {Barman}, \& {Hauschildt}}]{Husser2013}
{Husser}, T.~O., {Wende-von Berg}, S., {Dreizler}, S., {et~al.} 2013, \aap, 553, A6

\bibitem[{{Iyer} \& {Line}(2020)}]{Iyer2020}
{Iyer}, A.~R., \& {Line}, M.~R. 2020, \apj, 889, 78

\bibitem[{{Jakobsen} {et~al.}(2022){Jakobsen}, {Ferruit}, {Alves de Oliveira}, {Arribas}, {Bagnasco}, {Barho}, {Beck}, {Birkmann}, {B{\"o}ker}, {Bunker}, {Charlot}, {de Jong}, {de Marchi}, {Ehrenwinkler}, {Falcolini}, {Fels}, {Franx}, {Franz}, {Funke}, {Giardino}, {Gnata}, {Holota}, {Honnen}, {Jensen}, {Jentsch}, {Johnson}, {Jollet}, {Karl}, {Kling}, {K{\"o}hler}, {Kolm}, {Kumari}, {Lander}, {Lemke}, {L{\'o}pez-Caniego}, {L{\"u}tzgendorf}, {Maiolino}, {Manjavacas}, {Marston}, {Maschmann}, {Maurer}, {Messerschmidt}, {Moseley}, {Mosner}, {Mott}, {Muzerolle}, {Pirzkal}, {Pittet}, {Plitzke}, {Posselt}, {Rapp}, {Rauscher}, {Rawle}, {Rix}, {R{\"o}del}, {Rumler}, {Sabbi}, {Salvignol}, {Schmid}, {Sirianni}, {Smith}, {Strada}, {te Plate}, {Valenti}, {Wettemann}, {Wiehe}, {Wiesmayer}, {Willott}, {Wright}, {Zeidler}, \& {Zincke}}]{Jakobsen2022}
{Jakobsen}, P., {Ferruit}, P., {Alves de Oliveira}, C., {et~al.} 2022, \aap, 661, A80

\bibitem[{{JWST Transiting Exoplanet Community Early Release Science Team} {et~al.}(2023){JWST Transiting Exoplanet Community Early Release Science Team}, {Ahrer}, {Alderson}, {Batalha}, {Batalha}, {Bean}, {Beatty}, {Bell}, {Benneke}, {Berta-Thompson}, {Carter}, {Crossfield}, {Espinoza}, {Feinstein}, {Fortney}, {Gibson}, {Goyal}, {Kempton}, {Kirk}, {Kreidberg}, {L{\'o}pez-Morales}, {Line}, {Lothringer}, {Moran}, {Mukherjee}, {Ohno}, {Parmentier}, {Piaulet}, {Rustamkulov}, {Schlawin}, {Sing}, {Stevenson}, {Wakeford}, {Allen}, {Birkmann}, {Brande}, {Crouzet}, {Cubillos}, {Damiano}, {D{\'e}sert}, {Gao}, {Harrington}, {Hu}, {Kendrew}, {Knutson}, {Lagage}, {Leconte}, {Lendl}, {MacDonald}, {May}, {Miguel}, {Molaverdikhani}, {Moses}, {Murray}, {Nehring}, {Nikolov}, {Petit dit de la Roche}, {Radica}, {Roy}, {Stassun}, {Taylor}, {Waalkes}, {Wachiraphan}, {Welbanks}, {Wheatley}, {Aggarwal}, {Alam}, {Banerjee}, {Barstow}, {Blecic}, {Casewell}, {Changeat}, {Chubb}, {Col{\'o}n}, {Coulombe}, {Daylan}, {de Val-Borro},
  {Decin}, {Dos Santos}, {Flagg}, {France}, {Fu}, {Garc{\'\i}a Mu{\~n}oz}, {Gizis}, {Glidden}, {Grant}, {Heng}, {Henning}, {Hong}, {Inglis}, {Iro}, {Kataria}, {Komacek}, {Krick}, {Lee}, {Lewis}, {Lillo-Box}, {Lustig-Yaeger}, {Mancini}, {Mandell}, {Mansfield}, {Marley}, {Mikal-Evans}, {Morello}, {Nixon}, {Ortiz Ceballos}, {Piette}, {Powell}, {Rackham}, {Ramos-Rosado}, {Rauscher}, {Redfield}, {Rogers}, {Roman}, {Roudier}, {Scarsdale}, {Shkolnik}, {Southworth}, {Spake}, {Steinrueck}, {Tan}, {Teske}, {Tremblin}, {Tsai}, {Tucker}, {Turner}, {Valenti}, {Venot}, {Waldmann}, {Wallack}, {Zhang}, \& {Zieba}}]{ERSTeam2023}
{JWST Transiting Exoplanet Community Early Release Science Team}, {Ahrer}, E.-M., {Alderson}, L., {et~al.} 2023, \nat, 614, 649

\bibitem[{{Karman} {et~al.}(2019){Karman}, {Gordon}, {van der Avoird}, {Baranov}, {Boulet}, {Drouin}, {Groenenboom}, {Gustafsson}, {Hartmann}, {Kurucz}, {Rothman}, {Sun}, {Sung}, {Thalman}, {Tran}, {Wishnow}, {Wordsworth}, {Vigasin}, {Volkamer}, \& {van der Zande}}]{Karman2019}
{Karman}, T., {Gordon}, I.~E., {van der Avoird}, A., {et~al.} 2019, \icarus, 328, 160

\bibitem[{{Kataria} {et~al.}(2016){Kataria}, {Sing}, {Lewis}, {Visscher}, {Showman}, {Fortney}, \& {Marley}}]{Kataria2016}
{Kataria}, T., {Sing}, D.~K., {Lewis}, N.~K., {et~al.} 2016, \apj, 821, 9

\bibitem[{{Keating} {et~al.}(2019){Keating}, {Cowan}, \& {Dang}}]{Keating2019}
{Keating}, D., {Cowan}, N.~B., \& {Dang}, L. 2019, Nature Astronomy, 3, 1092

\bibitem[{{Keating} {et~al.}(2020){Keating}, {Stevenson}, {Cowan}, {Rauscher}, {Bean}, {Bell}, {Dang}, {Deming}, {D{\'e}sert}, {Feng}, {Fortney}, {Kataria}, {Kempton}, {Lewis}, {Line}, {Mansfield}, {May}, {Morley}, \& {Showman}}]{Keating2020}
{Keating}, D., {Stevenson}, K.~B., {Cowan}, N.~B., {et~al.} 2020, \aj, 159, 225

\bibitem[{{Kendrew} {et~al.}(2015){Kendrew}, {Scheithauer}, {Bouchet}, {Amiaux}, {Azzollini}, {Bouwman}, {Chen}, {Dubreuil}, {Fischer}, {Glasse}, {Greene}, {Lagage}, {Lahuis}, {Ronayette}, {Wright}, \& {Wright}}]{Kendrew2015}
{Kendrew}, S., {Scheithauer}, S., {Bouchet}, P., {et~al.} 2015, \pasp, 127, 623

\bibitem[{{Knutson} {et~al.}(2007){Knutson}, {Charbonneau}, {Allen}, {Fortney}, {Agol}, {Cowan}, {Showman}, {Cooper}, \& {Megeath}}]{KnutsonEtal2007natHD189733b}
{Knutson}, H.~A., {Charbonneau}, D., {Allen}, L.~E., {et~al.} 2007, \nat, 447, 183

\bibitem[{{Krick} {et~al.}(2016){Krick}, {Ingalls}, {Carey}, {von Braun}, {Kane}, {Ciardi}, {Plavchan}, {Wong}, \& {Lowrance}}]{Krick2016}
{Krick}, J.~E., {Ingalls}, J., {Carey}, S., {et~al.} 2016, \apj, 824, 27

\bibitem[{{Lee} {et~al.}(2023){Lee}, {Tsai}, {Hammond}, \& {Tan}}]{Lee2023}
{Lee}, E. K.~H., {Tsai}, S.-M., {Hammond}, M., \& {Tan}, X. 2023, \aap, 672, A110

\bibitem[{{Lewis} {et~al.}(2017){Lewis}, {Clampin}, {Mountain}, {Seager}, {Stevenson}, \& {Valenti}}]{Lewis_jwst.prop.1353L}
{Lewis}, N., {Clampin}, M., {Mountain}, M., {et~al.} 2017, {Transit and Eclipse Spectroscopy of a Hot Jupiter}, JWST Proposal. Cycle 1, ID. \#1353

\bibitem[{{Lewis} {et~al.}(2025){Lewis}, {Mullens}, {Wakeford}, {Batalha}, {Alderson}, {Glidden}, {Sotzen}, {Goyal}, {Challener}, {Valenti}, {Espinoza}, \& {et al.}}]{lewis2025_w17transit}
{Lewis}, N.~K., {Mullens}, E., {Wakeford}, H.~R., {et~al.} 2025, AJ, in prep

\bibitem[{{Li} {et~al.}(2015){Li}, {Gordon}, {Rothman}, {Tan}, {Hu}, {Kassi}, {Campargue}, \& {Medvedev}}]{Li2015}
{Li}, G., {Gordon}, I.~E., {Rothman}, L.~S., {et~al.} 2015, \apjs, 216, 15

\bibitem[{{Lim} {et~al.}(2023){Lim}, {Benneke}, {Doyon}, {MacDonald}, {Piaulet}, {Artigau}, {Coulombe}, {Radica}, {L'Heureux}, {Albert}, {Rackham}, {de Wit}, {Salhi}, {Roy}, {Flagg}, {Fournier-Tondreau}, {Taylor}, {Cook}, {Lafreni{\`e}re}, {Cowan}, {Kaltenegger}, {Rowe}, {Espinoza}, {Dang}, \& {Darveau-Bernier}}]{Lim2023}
{Lim}, O., {Benneke}, B., {Doyon}, R., {et~al.} 2023, \apjl, 955, L22

\bibitem[{{Limbach} {et~al.}(2022){Limbach}, {Vanderburg}, {Stevenson}, {Blouin}, {Morley}, {Lustig-Yaeger}, {Soares-Furtado}, \& {Janson}}]{Limbach2022}
{Limbach}, M.~A., {Vanderburg}, A., {Stevenson}, K.~B., {et~al.} 2022, \mnras, 517, 2622

\bibitem[{{Limbach} {et~al.}(2024){Limbach}, {Vanderburg}, {Venner}, {Blouin}, {Stevenson}, {MacDonald}, {Jenkins}, {Bowens-Rubin}, {Soares-Furtado}, {Morley}, {Janson}, {Debes}, {Xu}, {Kleisioti}, {Kenworthy}, {Butler}, {Crane}, {Osip}, {Shectman}, \& {Teske}}]{Limbach2024}
{Limbach}, M.~A., {Vanderburg}, A., {Venner}, A., {et~al.} 2024, \apjl, 973, L11

\bibitem[{{Line} {et~al.}(2015){Line}, {Teske}, {Burningham}, {Fortney}, \& {Marley}}]{Line2015}
{Line}, M.~R., {Teske}, J., {Burningham}, B., {Fortney}, J.~J., \& {Marley}, M.~S. 2015, \apj, 807, 183

\bibitem[{{L{\'o}pez-Morales} \& {Seager}(2007)}]{LopezMorales2007}
{L{\'o}pez-Morales}, M., \& {Seager}, S. 2007, \apjl, 667, L191

\bibitem[{{Louie} {et~al.}(2024){Louie}, {Mullens}, {Alderson}, {Glidden}, {Lewis}, {Wakeford}, {Batalha}, {Col{\'o}n}, {Gressier}, {Long}, {Radica}, {Espinoza}, {Goyal}, {MacDonald}, {May}, {Seager}, {Stevenson}, {Valenti}, {Allen}, {Ca{\~n}as}, {Challener}, {Grant}, {Huang}, {Lin}, {Valentine}, {Clampin}, {Perrin}, {Pueyo}, {van der Marel}, \& {Mountain}}]{Louie2024}
{Louie}, D.~R., {Mullens}, E., {Alderson}, L., {et~al.} 2024, arXiv e-prints, arXiv:2412.03675

\bibitem[{{Lustig-Yaeger} {et~al.}(2021){Lustig-Yaeger}, {Stevenson}, {Mayorga}, {Showalter Sotzen}, {May}, {Izenberg}, \& {Mandt}}]{Lustig-Yaeger2021}
{Lustig-Yaeger}, J., {Stevenson}, K.~B., {Mayorga}, L.~C., {et~al.} 2021, \apjl, 921, L4

\bibitem[{{Lustig-Yaeger} {et~al.}(2023){Lustig-Yaeger}, {Fu}, {May}, {Ortiz Ceballos}, {Moran}, {Peacock}, {Stevenson}, {L{\'o}pez-Morales}, {MacDonald}, {Mayorga}, {Sing}, {Sotzen}, {Valenti}, {Adams}, {Alam}, {Batalha}, {Bennett}, {Gonzalez-Quiles}, {Kirk}, {Kruse}, {Lothringer}, {Rustamkulov}, \& {Wakeford}}]{FuLustig2023}
{Lustig-Yaeger}, J., {Fu}, G., {May}, E.~M., {et~al.} 2023, arXiv e-prints, arXiv:2301.04191

\bibitem[{MacDonald(2023)}]{MacDonald2023}
MacDonald, R.~J. 2023, Journal of Open Source Software, 8, 4873

\bibitem[{{MacDonald} \& {Madhusudhan}(2017)}]{MacDonald2017}
{MacDonald}, R.~J., \& {Madhusudhan}, N. 2017, \mnras, 469, 1979

\bibitem[{{Mandell} {et~al.}(2013){Mandell}, {Haynes}, {Sinukoff}, {Madhusudhan}, {Burrows}, \& {Deming}}]{Mandell2013}
{Mandell}, A.~M., {Haynes}, K., {Sinukoff}, E., {et~al.} 2013, \apj, 779, 128

\bibitem[{{Mandell} {et~al.}(2022){Mandell}, {Lustig-Yaeger}, {Stevenson}, \& {Staguhn}}]{Mandell2022}
{Mandell}, A.~M., {Lustig-Yaeger}, J., {Stevenson}, K.~B., \& {Staguhn}, J. 2022, \aj, 164, 176

\bibitem[{{Matthews} {et~al.}(2024){Matthews}, {Watson}, {de Mooij}, {Marsh}, {Brogi}, {Merritt}, {Smith}, \& {Steeghs}}]{Matthews2024}
{Matthews}, S.~M., {Watson}, C.~A., {de Mooij}, E.~J.~W., {et~al.} 2024, \mnras, 531, 3800

\bibitem[{{May} {et~al.}(2023){May}, {MacDonald}, {Bennett}, {Moran}, {Wakeford}, {Peacock}, {Lustig-Yaeger}, {Highland}, {Stevenson}, {Sing}, {Mayorga}, {Batalha}, {Kirk}, {L{\'o}pez-Morales}, {Valenti}, {Alam}, {Alderson}, {Fu}, {Gonzalez-Quiles}, {Lothringer}, {Rustamkulov}, \& {Sotzen}}]{May2023}
{May}, E.~M., {MacDonald}, R.~J., {Bennett}, K.~A., {et~al.} 2023, \apjl, 959, L9

\bibitem[{{Mayorga} {et~al.}(2023){Mayorga}, {Lustig-Yaeger}, {Stevenson}, {Consortium On Habitability}, \& {Atmospheres Of M-Dwarf Planets (Champs)}}]{Mayorga2023}
{Mayorga}, L.~C., {Lustig-Yaeger}, J., {Stevenson}, K.~B., {Consortium On Habitability}, \& {Atmospheres Of M-Dwarf Planets (Champs)}. 2023, \apj, 956, 74

\bibitem[{{Mettler} {et~al.}(2024){Mettler}, {Konrad}, {Quanz}, \& {Helled}}]{Mettler2024}
{Mettler}, J.-N., {Konrad}, B.~S., {Quanz}, S.~P., \& {Helled}, R. 2024, \apj, 963, 24

\bibitem[{{Mikal-Evans} {et~al.}(2023){Mikal-Evans}, {Sing}, {Dong}, {Foreman-Mackey}, {Kataria}, {Barstow}, {Goyal}, {Lewis}, {Lothringer}, {Mayne}, {Wakeford}, {Christie}, \& {Rustamkulov}}]{MikalEvans2023}
{Mikal-Evans}, T., {Sing}, D.~K., {Dong}, J., {et~al.} 2023, \apjl, 943, L17

\bibitem[{{Molli{\`e}re} {et~al.}(2019){Molli{\`e}re}, {Wardenier}, {van Boekel}, {Henning}, {Molaverdikhani}, \& {Snellen}}]{Molliere2019}
{Molli{\`e}re}, P., {Wardenier}, J.~P., {van Boekel}, R., {et~al.} 2019, \aap, 627, A67

\bibitem[{{Moran} {et~al.}(2023){Moran}, {Stevenson}, {Sing}, {MacDonald}, {Kirk}, {Lustig-Yaeger}, {Peacock}, {Mayorga}, {Bennett}, {L{\'o}pez-Morales}, {May}, {Rustamkulov}, {Valenti}, {Adams Redai}, {Alam}, {Batalha}, {Fu}, {Gonzalez-Quiles}, {Highland}, {Kruse}, {Lothringer}, {Ortiz Ceballos}, {Sotzen}, \& {Wakeford}}]{Moran2023}
{Moran}, S.~E., {Stevenson}, K.~B., {Sing}, D.~K., {et~al.} 2023, \apjl, 948, L11

\bibitem[{{Morris} {et~al.}(2022){Morris}, {Heng}, {Jones}, {Piaulet}, {Demory}, {Kitzmann}, \& {Jens Hoeijmakers}}]{Morris2022}
{Morris}, B.~M., {Heng}, K., {Jones}, K., {et~al.} 2022, \aap, 660, A123

\bibitem[{{Moses} {et~al.}(2011){Moses}, {Visscher}, {Fortney}, {Showman}, {Lewis}, {Griffith}, {Klippenstein}, {Shabram}, {Friedson}, {Marley}, \& {Freedman}}]{Moses2011}
{Moses}, J.~I., {Visscher}, C., {Fortney}, J.~J., {et~al.} 2011, \apj, 737, 15

\bibitem[{{Moses} {et~al.}(2013){Moses}, {Line}, {Visscher}, {Richardson}, {Nettelmann}, {Fortney}, {Barman}, {Stevenson}, \& {Madhusudhan}}]{Moses2013b}
{Moses}, J.~I., {Line}, M.~R., {Visscher}, C., {et~al.} 2013, \apj, 777, 34

\bibitem[{{Mraz} {et~al.}(2024){Mraz}, {Darveau-Bernier}, {Boucher}, {Cowan}, {Lafreni{\`e}re}, \& {Cadieux}}]{Mraz2024}
{Mraz}, G., {Darveau-Bernier}, A., {Boucher}, A., {et~al.} 2024, \apjl, 975, L42

\bibitem[{{Mullens} {et~al.}(2024){Mullens}, {Lewis}, \& {MacDonald}}]{Mullens2024}
{Mullens}, E., {Lewis}, N.~K., \& {MacDonald}, R.~J. 2024, \apj, 977, 105

\bibitem[{{Pelletier} {et~al.}(2021){Pelletier}, {Benneke}, {Darveau-Bernier}, {Boucher}, {Cook}, {Piaulet}, {Coulombe}, {Artigau}, {Lafreni{\`e}re}, {Delisle}, {Allart}, {Doyon}, {Donati}, {Fouqu{\'e}}, {Moutou}, {Cadieux}, {Delfosse}, {H{\'e}brard}, {Martins}, {Martioli}, \& {Vandal}}]{Pelletier2021}
{Pelletier}, S., {Benneke}, B., {Darveau-Bernier}, A., {et~al.} 2021, \aj, 162, 73

\bibitem[{{Polyansky} {et~al.}(2018){Polyansky}, {Kyuberis}, {Zobov}, {Tennyson}, {Yurchenko}, \& {Lodi}}]{Polyansky2018}
{Polyansky}, O.~L., {Kyuberis}, A.~A., {Zobov}, N.~F., {et~al.} 2018, \mnras, 480, 2597

\bibitem[{{Powell} {et~al.}(2024){Powell}, {Feinstein}, {Lee}, {Zhang}, {Tsai}, {Taylor}, {Kirk}, {Bell}, {Barstow}, {Gao}, {Bean}, {Blecic}, {Chubb}, {Crossfield}, {Jordan}, {Kitzmann}, {Moran}, {Morello}, {Moses}, {Welbanks}, {Yang}, {Zhang}, {Ahrer}, {Bello-Arufe}, {Brande}, {Casewell}, {Crouzet}, {Cubillos}, {Demory}, {Dyrek}, {Flagg}, {Hu}, {Inglis}, {Jones}, {Kreidberg}, {L{\'o}pez-Morales}, {Lagage}, {Meier Vald{\'e}s}, {Miguel}, {Parmentier}, {Piette}, {Rackham}, {Radica}, {Redfield}, {Stevenson}, {Wakeford}, {Aggarwal}, {Alam}, {Batalha}, {Batalha}, {Benneke}, {Berta-Thompson}, {Brady}, {Caceres}, {Carter}, {D{\'e}sert}, {Harrington}, {Iro}, {Line}, {Lothringer}, {MacDonald}, {Mancini}, {Molaverdikhani}, {Mukherjee}, {Nixon}, {Oza}, {Palle}, {Rustamkulov}, {Sing}, {Steinrueck}, {Venot}, {Wheatley}, \& {Yurchenko}}]{Powell2024}
{Powell}, D., {Feinstein}, A.~D., {Lee}, E. K.~H., {et~al.} 2024, \nat, 626, 979

\bibitem[{Rackham {et~al.}(2017)Rackham, Espinoza, Apai, L{\'{o}}pez-Morales, Jord{\'{a}}n, Osip, Lewis, Rodler, Fraine, Morley, \& Fortney}]{Rackham2017}
Rackham, B., Espinoza, N., Apai, D., {et~al.} 2017, The Astrophysical Journal, 834, 151

\bibitem[{{Rackham} \& {de Wit}(2024)}]{Rackham2024}
{Rackham}, B.~V., \& {de Wit}, J. 2024, \aj, 168, 82

\bibitem[{{Rackham} {et~al.}(2023){Rackham}, {Espinoza}, {Berdyugina}, {Korhonen}, {MacDonald}, {Montet}, {Morris}, {Oshagh}, {Shapiro}, {Unruh}, {Quintana}, {Zellem}, {Apai}, {Barclay}, {Barstow}, {Bruno}, {Carone}, {Casewell}, {Cegla}, {Criscuoli}, {Fischer}, {Fournier}, {Giampapa}, {Giles}, {Iyer}, {Kopp}, {Kostogryz}, {Krivova}, {Mallonn}, {McGruder}, {Molaverdikhani}, {Newton}, {Panja}, {Peacock}, {Reardon}, {Roettenbacher}, {Scandariato}, {Solanki}, {Stassun}, {Steiner}, {Stevenson}, {Tregloan-Reed}, {Valio}, {Wedemeyer}, {Welbanks}, {Yu}, {Alam}, {Davenport}, {Deming}, {Dong}, {Ducrot}, {Fisher}, {Gilbert}, {Kostov}, {L{\'o}pez-Morales}, {Line}, {Mo{\v{c}}nik}, {Mullally}, {Paudel}, {Ribas}, \& {Valenti}}]{Rackham2023}
{Rackham}, B.~V., {Espinoza}, N., {Berdyugina}, S.~V., {et~al.} 2023, RAS Techniques and Instruments, 2, 148

\bibitem[{{Radica} {et~al.}(2025){Radica}, {Piaulet-Ghorayeb}, {Taylor}, {Coulombe}, {Benneke}, {Albert}, {Artigau}, {Cowan}, {Doyon}, {Lafreni{\`e}re}, {L'Heureux}, \& {Lim}}]{Radica2025}
{Radica}, M., {Piaulet-Ghorayeb}, C., {Taylor}, J., {et~al.} 2025, \apjl, 979, L5

\bibitem[{{Rugheimer} {et~al.}(2013){Rugheimer}, {Kaltenegger}, {Zsom}, {Segura}, \& {Sasselov}}]{Rugheimer2013}
{Rugheimer}, S., {Kaltenegger}, L., {Zsom}, A., {Segura}, A., \& {Sasselov}, D. 2013, Astrobiology, 13, 251

\bibitem[{{Rustamkulov} {et~al.}(2023){Rustamkulov}, {Sing}, {Mukherjee}, {May}, {Kirk}, {Schlawin}, {Line}, {Piaulet}, {Carter}, {Batalha}, {Goyal}, {L{\'o}pez-Morales}, {Lothringer}, {MacDonald}, {Moran}, {Stevenson}, {Wakeford}, {Espinoza}, {Bean}, {Batalha}, {Benneke}, {Berta-Thompson}, {Crossfield}, {Gao}, {Kreidberg}, {Powell}, {Cubillos}, {Gibson}, {Leconte}, {Molaverdikhani}, {Nikolov}, {Parmentier}, {Roy}, {Taylor}, {Turner}, {Wheatley}, {Aggarwal}, {Ahrer}, {Alam}, {Alderson}, {Allen}, {Banerjee}, {Barat}, {Barrado}, {Barstow}, {Bell}, {Blecic}, {Brande}, {Casewell}, {Changeat}, {Chubb}, {Crouzet}, {Daylan}, {Decin}, {D{\'e}sert}, {Mikal-Evans}, {Feinstein}, {Flagg}, {Fortney}, {Harrington}, {Heng}, {Hong}, {Hu}, {Iro}, {Kataria}, {Kempton}, {Krick}, {Lendl}, {Lillo-Box}, {Louca}, {Lustig-Yaeger}, {Mancini}, {Mansfield}, {Mayne}, {Miguel}, {Morello}, {Ohno}, {Palle}, {Petit dit de la Roche}, {Rackham}, {Radica}, {Ramos-Rosado}, {Redfield}, {Rogers}, {Shkolnik}, {Southworth}, {Teske}, {Tremblin},
  {Tucker}, {Venot}, {Waalkes}, {Welbanks}, {Zhang}, \& {Zieba}}]{Rustamkulov2023}
{Rustamkulov}, Z., {Sing}, D.~K., {Mukherjee}, S., {et~al.} 2023, Nature, 614, 659

\bibitem[{{Schwartz} \& {Cowan}(2015)}]{Schwartz2015}
{Schwartz}, J.~C., \& {Cowan}, N.~B. 2015, \mnras, 449, 4192

\bibitem[{{Sedaghati} {et~al.}(2016){Sedaghati}, {Boffin}, {Je{\v{r}}abkov{\'a}}, {Garc{\'\i}a Mu{\~n}oz}, {Grenfell}, {Smette}, {Ivanov}, {Csizmadia}, {Cabrera}, {Kabath}, {Rocchetto}, \& {Rauer}}]{Sedaghati2016}
{Sedaghati}, E., {Boffin}, H.~M.~J., {Je{\v{r}}abkov{\'a}}, T., {et~al.} 2016, \aap, 596, A47

\bibitem[{{Showman} {et~al.}(2020){Showman}, {Tan}, \& {Parmentier}}]{Showman2020}
{Showman}, A.~P., {Tan}, X., \& {Parmentier}, V. 2020, \ssr, 216, 139

\bibitem[{{Sikora} {et~al.}(2025){Sikora}, {Rowe}, {Splinter}, {Barat}, {Dang}, {Cowan}, {Barclay}, {Col{\'o}n}, {D{\'e}sert}, {Kane}, {Llama}, {Shivkumar}, {Stassun}, \& {Quintana}}]{Sikora2025}
{Sikora}, J.~T., {Rowe}, J.~F., {Splinter}, J., {et~al.} 2025, \aj, 170, 105

\bibitem[{{Sing} {et~al.}(2016){Sing}, {Fortney}, {Nikolov}, {Wakeford}, {Kataria}, {Evans}, {Aigrain}, {Ballester}, {Burrows}, {Deming}, {D{\'e}sert}, {Gibson}, {Henry}, {Huitson}, {Knutson}, {Etangs}, {Pont}, {Showman}, {Vidal-Madjar}, {Williamson}, \& {Wilson}}]{Sing2016}
{Sing}, D.~K., {Fortney}, J.~J., {Nikolov}, N., {et~al.} 2016, \nat, 529, 59

\bibitem[{{Smitha} {et~al.}(2025){Smitha}, {Shapiro}, {Witzke}, {Kostogryz}, {Unruh}, {Bhatia}, {Cameron}, {Seager}, \& {Solanki}}]{Smitha2025}
{Smitha}, H.~N., {Shapiro}, A.~I., {Witzke}, V., {et~al.} 2025, \apjl, 978, L13

\bibitem[{{Sotzen} {et~al.}(2025){Sotzen}, {Lustig-Yaeger}, {Stevenson}, {Tsai}, \& {et al.}}]{hatp26b_pie}
{Sotzen}, K.~S., {Lustig-Yaeger}, J., {Stevenson}, K.~B., {Tsai}, S.~M., \& {et al.} 2025, RNAAS, in prep

\bibitem[{{Speagle}(2020)}]{Speagle2020}
{Speagle}, J.~S. 2020, \mnras, 493, 3132

\bibitem[{{Spiegel} \& {Burrows}(2010)}]{Spiegel2010}
{Spiegel}, D.~S., \& {Burrows}, A. 2010, \apj, 722, 871

\bibitem[{{Splinter} {et~al.}(2025){Splinter}, {Coulombe}, {Frazier}, {Cowan}, {Rauscher}, {Dang}, {Radica}, {Collins}, {Pelletier}, {Allart}, {MacDonald}, {Lafreni{\`e}re}, {Albert}, {Benneke}, {Doyon}, {Jayawardhana}, {Johnstone}, {Krishnamurthy}, {Piaulet-Ghorayeb}, {Kaltnegger}, {Meyer}, {Taylor}, \& {Turner}}]{Splinter2025}
{Splinter}, J., {Coulombe}, L.-P., {Frazier}, R.~C., {et~al.} 2025, arXiv e-prints, arXiv:2509.09760

\bibitem[{{Stassun} {et~al.}(2017){Stassun}, {Collins}, \& {Gaudi}}]{Stassun2017}
{Stassun}, K.~G., {Collins}, K.~A., \& {Gaudi}, B.~S. 2017, \aj, 153, 136

\bibitem[{{Stevenson}(2020)}]{Stevenson2020}
{Stevenson}, K.~B. 2020, \apjl, 898, L35

\bibitem[{{Stevenson} {et~al.}(2014){Stevenson}, {Desert}, {Line}, {Bean}, {Fortney}, {Showman}, {Kataria}, {Kreidberg}, {McCullough}, {Henry}, {Charbonneau}, {Burrows}, {Seager}, {Madhusudhan}, {Williamson}, \& {Homeier}}]{Stevenson2014c}
{Stevenson}, K.~B., {Desert}, J.-M., {Line}, M.~R., {et~al.} 2014, Science, 346, 838

\bibitem[{{STScI Development Team}(2013)}]{pysynphot}
{STScI Development Team}. 2013, {pysynphot: Synthetic photometry software package}, Astrophysics Source Code Library, record ascl:1303.023, ascl:1303.023

\bibitem[{{STScI Development Team}(2018)}]{STScI2018}
---. 2018, {synphot: Synthetic photometry using Astropy}, Astrophysics Source Code Library, record ascl:1811.001

\bibitem[{{Tinetti} {et~al.}(2020){Tinetti}, {Eccleston}, {Lueftinger}, {Pilbratt}, \& {Puig}}]{ARIEL2020}
{Tinetti}, G., {Eccleston}, P., {Lueftinger}, T., {Pilbratt}, G., \& {Puig}, L. 2020, in European Planetary Science Congress, EPSC2020--696

\bibitem[{{Tsai} {et~al.}(2021){Tsai}, {Malik}, {Kitzmann}, {Lyons}, {Fateev}, {Lee}, \& {Heng}}]{Tsai2021}
{Tsai}, S.-M., {Malik}, M., {Kitzmann}, D., {et~al.} 2021, \apj, 923, 264

\bibitem[{{Tsai} {et~al.}(2023){Tsai}, {Moses}, {Powell}, \& {Lee}}]{Tsai2023c}
{Tsai}, S.-M., {Moses}, J.~I., {Powell}, D., \& {Lee}, E. K.~H. 2023, \apjl, 959, L30

\bibitem[{{Tsai} {et~al.}(2023b){Tsai}, {Lee}, {Powell}, {Gao}, {Zhang}, {Moses}, {H{\'e}brard}, {Venot}, {Parmentier}, {Jordan}, {Hu}, {Alam}, {Alderson}, {Batalha}, {Bean}, {Benneke}, {Bierson}, {Brady}, {Carone}, {Carter}, {Chubb}, {Inglis}, {Leconte}, {Line}, {L{\'o}pez-Morales}, {Miguel}, {Molaverdikhani}, {Rustamkulov}, {Sing}, {Stevenson}, {Wakeford}, {Yang}, {Aggarwal}, {Baeyens}, {Barat}, {de Val-Borro}, {Daylan}, {Fortney}, {France}, {Goyal}, {Grant}, {Kirk}, {Kreidberg}, {Louca}, {Moran}, {Mukherjee}, {Nasedkin}, {Ohno}, {Rackham}, {Redfield}, {Taylor}, {Tremblin}, {Visscher}, {Wallack}, {Welbanks}, {Youngblood}, {Ahrer}, {Batalha}, {Behr}, {Berta-Thompson}, {Blecic}, {Casewell}, {Crossfield}, {Crouzet}, {Cubillos}, {Decin}, {D{\'e}sert}, {Feinstein}, {Gibson}, {Harrington}, {Heng}, {Henning}, {Kempton}, {Krick}, {Lagage}, {Lendl}, {Lothringer}, {Mansfield}, {Mayne}, {Mikal-Evans}, {Palle}, {Schlawin}, {Shorttle}, {Wheatley}, \& {Yurchenko}}]{Tsai2023a}
{Tsai}, S.-M., {Lee}, E. K.~H., {Powell}, D., {et~al.} 2023b, \nat, 617, 483

\bibitem[{{Tsai} {et~al.}(2024){Tsai}, {Parmentier}, {Mendon{\c{c}}a}, {Tan}, {Deitrick}, {Hammond}, {Savel}, {Zhang}, {Pierrehumbert}, \& {Schwieterman}}]{Tsai2024}
{Tsai}, S.-M., {Parmentier}, V., {Mendon{\c{c}}a}, J.~M., {et~al.} 2024, \apj, 963, 41

\bibitem[{{Underwood} {et~al.}(2016){Underwood}, {Tennyson}, {Yurchenko}, {Huang}, {Schwenke}, {Lee}, {Clausen}, \& {Fateev}}]{Underwood2016}
{Underwood}, D.~S., {Tennyson}, J., {Yurchenko}, S.~N., {et~al.} 2016, \mnras, 459, 3890

\bibitem[{{Valentine} {et~al.}(2024){Valentine}, {Wakeford}, {Challener}, {Batalha}, {Lewis}, {Grant}, {Mullens}, {Alderson}, {Goyal}, {MacDonald}, {May}, {Seager}, {Stevenson}, {Valenti}, {Allen}, {Espinoza}, {Glidden}, {Gressier}, {Huang}, {Lin}, {Long}, {Louie}, {Clampin}, {Perrin}, {van der Marel}, \& {Mountain}}]{Valentine2024}
{Valentine}, D., {Wakeford}, H.~R., {Challener}, R.~C., {et~al.} 2024, \aj, 168, 123

\bibitem[{{Wakeford} {et~al.}(2015){Wakeford}, {Sing}, {Evans}, {Deming}, \& {Mandell}}]{Wakeford2016}
{Wakeford}, H.~R., {Sing}, D.~K., {Evans}, T., {Deming}, D., \& {Mandell}, A. 2015, \aap, 573, 13

\bibitem[{{Wakeford} {et~al.}(2018){Wakeford}, {Sing}, {Deming}, {Lewis}, {Goyal}, {Wilson}, {Barstow}, {Kataria}, {Drummond}, {Evans}, {Carter}, {Nikolov}, {Knutson}, {Ballester}, \& {Mandell}}]{Wakeford2018}
{Wakeford}, H.~R., {Sing}, D.~K., {Deming}, D., {et~al.} 2018, \aj, 155, 29

\bibitem[{{Wakeford} {et~al.}(2019){Wakeford}, {Lewis}, {Fowler}, {Bruno}, {Wilson}, {Moran}, {Valenti}, {Batalha}, {Filippazzo}, {Bourrier}, {H{\"o}rst}, {Lederer}, \& {de Wit}}]{Wakeford2019}
{Wakeford}, H.~R., {Lewis}, N.~K., {Fowler}, J., {et~al.} 2019, \aj, 157, 11

\bibitem[{{Wakeford} {et~al.}(2025){Wakeford}, {Challener}, {Lewis}, {Goyal}, {Batalha}, {Alderson}, {Gressier}, {Mullens}, {Sotzen}, {Valenti}, {Espinoza}, \& {et al.}}]{wakeford2025_w17emission}
{Wakeford}, H.~R., {Challener}, R., {Lewis}, N.~K., {et~al.} 2025, AJ, in prep

\bibitem[{{Waldmann} {et~al.}(2015){Waldmann}, {Rocchetto}, {Tinetti}, {Barton}, {Yurchenko}, \& {Tennyson}}]{Waldmann2015}
{Waldmann}, I.~P., {Rocchetto}, M., {Tinetti}, G., {et~al.} 2015, \apj, 813, 13

\bibitem[{{Witzke} {et~al.}(2022){Witzke}, {Shapiro}, {Kostogryz}, {Cameron}, {Rackham}, {Seager}, {Solanki}, \& {Unruh}}]{Witzke2022}
{Witzke}, V., {Shapiro}, A.~I., {Kostogryz}, N.~M., {et~al.} 2022, \apjl, 941, L35

\bibitem[{{Wong} {et~al.}(2014){Wong}, {Knutson}, {Cowan}, {Lewis}, {Agol}, {Burrows}, {Deming}, {Fortney}, {Fulton}, {Langton}, {Laughlin}, \& {Showman}}]{Wong2014}
{Wong}, I., {Knutson}, H.~A., {Cowan}, N.~B., {et~al.} 2014, \apj, 794, 134

\bibitem[{{Wood} {et~al.}(2011){Wood}, {Maxted}, {Smalley}, \& {Iro}}]{Wood2011}
{Wood}, P.~L., {Maxted}, P.~F.~L., {Smalley}, B., \& {Iro}, N. 2011, \mnras, 412, 2376

\bibitem[{{Yang} {et~al.}(2024){Yang}, {Chen}, {Yan}, {Tan}, \& {Ji}}]{Yang2024}
{Yang}, Y., {Chen}, G., {Yan}, F., {Tan}, X., \& {Ji}, J. 2024, \apjl, 971, L8

\bibitem[{{Yurchenko} {et~al.}(2020){Yurchenko}, {Mellor}, {Freedman}, \& {Tennyson}}]{Yurchenko2020}
{Yurchenko}, S.~N., {Mellor}, T.~M., {Freedman}, R.~S., \& {Tennyson}, J. 2020, \mnras, 496, 5282

\bibitem[{{Yurchenko} {et~al.}(2024){Yurchenko}, {Owens}, {Kefala}, \& {Tennyson}}]{Yurchenko2024}
{Yurchenko}, S.~N., {Owens}, A., {Kefala}, K., \& {Tennyson}, J. 2024, \mnras, 528, 3719

\bibitem[{{Zahnle} {et~al.}(2009){Zahnle}, {Marley}, {Freedman}, {Lodders}, \& {Fortney}}]{Zahnle2009b}
{Zahnle}, K., {Marley}, M.~S., {Freedman}, R.~S., {Lodders}, K., \& {Fortney}, J.~J. 2009, Astrophy. J. Lett., 701, L20

\bibitem[{{Zamyatina} {et~al.}(2023){Zamyatina}, {H{\'e}brard}, {Drummond}, {Mayne}, {Manners}, {Christie}, {Tremblin}, {Sing}, \& {Kohary}}]{Zamyatina2023}
{Zamyatina}, M., {H{\'e}brard}, E., {Drummond}, B., {et~al.} 2023, \mnras, 519, 3129

\end{thebibliography}

\end{document}